\documentclass[11pt]{article}
\usepackage{amsfonts,latexsym}
\usepackage{amsmath}
\usepackage{amscd}
\usepackage{float,amsmath,amssymb,mathrsfs,bm,multirow,graphics}
\usepackage[dvips]{graphicx}

\newcommand{\nequation}{\setcounter{equation}{0}}
\renewcommand{\theequation}{\mbox{\arabic{section}.\arabic{equation}}}
\newcommand{\R}{{\Bbb R}}

\newcommand{\C}{{\Bbb C}}

\newcommand{\proofbegin}{\noindent{\it Proof.\,\,}}

\newtheorem{theorem}{Theorem}[section]

\newtheorem{assumption}[theorem]{Assumption}
\newtheorem{remark}[theorem]{Remark}

\newtheorem{figuretext}[theorem]{Figure}

\input epsf

\begin{document}

\begin{center}

{\LARGE \sc Explicit soliton asymptotics for the Korteweg-de Vries equation on the half-line\\} \vspace {7mm}  \noindent

{\large A. S. Fokas$^{a}$} and {\large J. Lenells$^{b}$}

\vskip.7cm

\hskip-.6cm
\begin{tabular}{c}
$\phantom{R^R}^{a}${\small Department of Applied Mathematics and Theoretical Physics, University of Cambridge,  }
\\ {\small Cambridge CB3 0WA, United Kingdom} \\
{\small E-mail: t.fokas@damtp.cam.ac.uk} \\
\\
$\phantom{R^R}^{b}${\small Institut f\"ur Angewandte Mathematik, Leibniz Universit\"at Hannover}\\ 
{\small Welfengarten 1, 30167 Hannover, Germany} \\
{\small E-mail: lenells@ifam.uni-hannover.de} \\
\\
\end{tabular}
\vskip.5cm
\end{center}

\begin{abstract} 
\noindent
There exists a distinctive class of physically significant evolution PDEs in one spatial dimension which can be treated analytically. A prototypical example of this class (which is referred to as integrable), is the Korteweg-de Vries equation. Integrable PDEs on the line can be analyzed by the so-called Inverse Scattering Transform (IST) method. A particularly powerful aspect of the IST is its ability to predict the large $t$ behavior of the solution. Namely, starting with initial data $u(x,0)$, IST implies that the solution $u(x,t)$ asymptotes to a collection of solitons as $t \to \infty$, $x/t = O(1)$; moreover the shapes and speeds of these solitons can be computed from $u(x,0)$ using only {\it linear} operations. One of the most important developments in this area has been the generalization of the IST from initial to initial-boundary value (IBV) problems formulated on the half-line. It can be shown that again $u(x,t)$ asymptotes into a collection of solitons, where now the shapes and the speeds of these solitons depend both on $u(x,0)$ and on the given boundary conditions at $x = 0$. A major complication of IBV problems is that the computation of the shapes and speeds of the solitons involves the solution of a {\it nonlinear} Volterra integral equation. However, for a certain class of boundary conditions, called linearizable, this complication can be bypassed and the relevant computation is as effective as in the case of the problem on the line. Here, after reviewing the general theory for KdV, we analyze three different types of linearizable boundary conditions. For these cases, the initial conditions are: (a) restrictions of one and two soliton solutions at $t = 0$; (b) profiles of certain exponential type; (c) box-shaped profiles. For each of these cases, by computing explicitly the shapes and the speeds of the asymptotic solitons, we elucidate the influence of the boundary.
 \end{abstract}



\tableofcontents

\section{Introduction} \nequation
Systems governed by nonlinear differential equations are of fundamental importance in all branches of science. Although our understanding of nonlinear equations in general is still very limited, the last few decades have seen some important developments in the ways we describe, model, and predict certain nonlinear phenomena. One of the most exciting developments in this direction has been the implementation of the Inverse Scattering Transform (IST) for integrable nonlinear evolution equations in $1 + 1$ dimensions and the associated emergence of soliton theory. A particularly powerful aspect of the IST is its ability to predict the large time asymptotics of the solution. Given an integrable evolution equation and initial data $u_0(x)$ defined for $x \in \R$, the IST formalism implies that for large $t$ the solution $u(x,t)$ will asymptote to a collection of solitons traveling at constant speeds. Moreover, these asymptotic solitons can be determined from the initial data: Each soliton corresponds to a zero of an analytic spectral function $a(k)$ defined in terms of $u_0(x)$ by means of purely linear operations. 

One of the most important developments of soliton theory has been the generalization of the IST formalism from initial value to initial-boundary value (IBV) problems \cite{F1997, F2002, F-I-S}. Such problems appear in many applications, where it is often more natural to assume that the space variable is defined only on part of the real axis. For example, in the context of water waves it is usually far easier to measure the elevation of the water's surface over time at a fixed position (say at $x = 0$), than to determine the surface for all values of $x$ at a specific instant of time. This leads naturally to the mathematical formulation of an IBV problem on the half-line $\{x > 0\}$ with a prescribed boundary condition at $x = 0$ and with vanishing initial conditions. 

For IBV problems, an important issue, both for mathematical and physical considerations, is the study of the effect of the boundary on the asymptotic behavior of the solution. The approach of \cite{F1997, F2002} yields an expression for the solution $u(x,t)$ of an IBV problem in terms of the solution of a Riemann-Hilbert (RH) problem. All the ingredients of this RH problem are defined in terms of the initial and boundary values of $u(x,t)$. 
The asymptotic behavior of $u(x,t)$ can be analyzed in an effective way by using this RH problem and by employing a nonlinear version of the steepest descent method \cite{D-Z}. Using this approach it can be shown \cite{FI92, FI94, FI96, Kamvissis} that, just like in the infinite line setting, the solution on the half-line will asymptote for large $t$ to a collection of solitons traveling at constant speeds. It should be emphasized that on the half-line the asymptotic solitons are not generated by zeros of the equivalent of $a(k)$ (which is specified by the initial conditions), but they are generated by the zeros of a certain function $d(k)$ whose definition involves the initial conditions as well as the boundary values at $x = 0$. Thus, the boundary can have a nontrivial effect on the asymptotics of the solution.

\subsection{Purpose of the paper}
The aspects of the half-line spectral theory mentioned above have been analyzed in general terms in several papers (see for example \cite{BFS2004, BFS2006, BK2003, BS2003, F2004, FI2004, FM1999, L-FGNLS, MK2006, P2005}), but very few explicit examples have been studied either analytically or numerically. This is partly due to the fact that the effective construction of the solutions of IBV problems involves a certain step which is {\it nonlinear}: Despite the fact that the solution $u(x,t)$ {\it can} be recovered from the solution of a RH problem formulated uniquely in terms of the values of $u(x,0)$ on the half-line $\{x> 0, t=0\}$ together with the values of $u(x,t)$ and its $x$-derivatives on the boundary $\{x = 0, t > 0\}$, this does not provide directly an effective formula for solving the problem. Indeed, for a given IBV problem only part of the boundary values are prescribed as boundary conditions. The remaining boundary values cannot be independently specified, but are determined by the analysis of the so-called global relation. This yields the unknown boundary values in terms of the known initial and boundary conditions via the solution of a {\it nonlinear} Volterra integral equation \cite{BFS2003, F2005, TF2008}.

Nevertheless, there exists a class of IBV problems for which the global relation can be solved explicitly. In this case all the ingredients needed for the formulation of the RH problem can be obtained as effectively as in the case of the initial value problem. These so-called linearizable IBV problems provide a class of problems for which the asymptotic behavior of the solution on the half-line can be analyzed with the same effectiveness as the corresponding problem on the infinite line. 

The purpose of the present paper is to provide the analysis of a number of concrete examples of linearizable IBV problems. The explicit formulas presented here elucidate the influence of the boundary on the asymptotic behavior of the solution. 

\subsection{The KdV with dominant surface tension}
Our approach for constructing explicit solutions will be illustrated using the following equation:
\begin{equation}\label{kdv}
  u_t + u_x - u_{xxx} + 6u u_x = 0.
\end{equation}  
This equation is the KdV equation in the case of dominant surface tension. Indeed, the Korteweg-de Vries equation for long one-dimensional, small-amplitude, gravity waves propagating in shallow water is
\begin{equation}\label{kdvphysical}
  \eta_\tau = \frac{3}{2}\sqrt{\frac{g}{h}} \left(\frac{1}{2}\eta^2 + \frac{2}{3}\alpha \eta + \frac{1}{3} \sigma \eta_{\xi\xi}\right)_\xi, \qquad \sigma = \frac{1}{3}h^3 - \frac{T h}{\rho g},
\end{equation}
where $\eta$ is the surface elevation of the wave above the equilibrium level $h$, $\alpha$ is a parameter related to the uniform motion of the liquid, $g$ is the gravitational constant, $T$ is the surface tension, and $\rho$ is the density cf. \cite{A-C}. In the presence of sufficiently large surface tension, the parameter $\sigma$ is negative, and equation (\ref{kdvphysical}) can be brought into the nondimensional form (\ref{kdv}).

The reason for considering equation (\ref{kdv}) instead of the standard KdV equation is that whereas the IBV for the standard KdV equation on the half-line is well-posed with one boundary condition at $x = 0$, equation (\ref{kdv}) is well-posed if two boundary conditions are specified at $x = 0$. For our purposes the latter situation is favorable since it gives more freedom in the choice of boundary data at $x = 0$. In particular, equation (\ref{kdv}) admits the following two classes of linearizable boundary conditions (see section \ref{linearizablesec}):
\begin{itemize}
\item[(a)] $u(0,t) = u_{xx}(0,t) = 0, \quad t > 0,$
\item[(b)] 
$u(0,t) = \chi, \quad u_{xx}(0,t) = \chi + 3\chi^2, \quad t > 0, \quad \text{for some $\chi \in \R$ with $\chi \neq 0$.}$
\end{itemize}
We will refer to boundary conditions of these two types as linearizable of type (a) or type (b), respectively.

\subsection{The examples}
We will consider three classes of examples: (i) Solitons (ii) IBV problems with exponential intitial profiles (iii) IBV problems with box-shaped initial profiles. These classes of examples allow us to analyze problems with both types of linearizable boundary conditions: the IBV problems with box-shaped intitial profiles are associated with linearizable boundary conditions of type (a), and the IBV problems with exponential intitial profiles are associated with linearizable boundary conditions of type (b).\footnote{For the soliton solutions we can solve everything explicitly and will not need the special machinery available for linearizable boundary conditions. Nonetheless, for some choices of the parameters the boundary conditions turn out to be linearizable of type (b), and in these cases we will implement the linearizable approach as a check on the formalism.}

We believe the study of explicit examples is of interest since: (a) it provides a nontrivial verification of the general formalism; (b) it reveals interesting properties that were previously unnoticed; (c) it makes more accessible the general rather abstract theory.

\subsubsection{Solitons} 
Equation (\ref{kdv}) formulated on the line, $-\infty < x < \infty$, admits solitons. We can construct an explicit solution to an IBV problem on the half-line by simply restricting such a soliton solution to $x > 0$ and choosing compatible boundary conditions at $x = 0$. We will consider the restrictions of one-solitons and two-solitons and also comment on the case of rational solitons. 

The IBV problems constructed in this manner are of course artificial, but they are important for pedagogical considerations: In this case the spectral functions and all the ingredients of the RH problem can be computed directly, thus the analysis of these problems provides an important check on the general formalism and also gives us insight into nontrivial examples. 

We will compute the function $d(k)$ whose zeros generate the solitons and verify that the general theory predicts the correct answer. 
Moreover, it turns out that for certain choices of the parameter values, these restricted soliton solutions satisfy a linearizable boundary condition at $x =0$. Hence, for these cases we can also apply the approach available for linearizable boundary conditions. In this approach the function $d(k)$ is replaced by another function $\Delta(k)$ constructed in terms of the initial data alone; the theory claims that the zeros of $\Delta(k)$ generate the asymptotic solitons. We again verify that the linearizable theory gives the correct result. 

\subsubsection{Exponential intitial profiles}
Consider the family of initial profiles $E_r(x)$ defined by
\begin{equation}\label{Erxdefintro}  
  E_r(x) = \frac{1}{3} e^{-r x} \left(r^2-1\right),\qquad r > 0, \, r \neq 1, \, \text{$r$ constant}.
\end{equation}
The initial and boundary conditions
\begin{equation}\label{expIBVintro}  
\begin{cases}
  u_r(x, 0) = E_r(x), \qquad x > 0, \\
  u_r(0,t) = \chi_r, \quad u_{rxx}(0,t) = \chi_r + 3\chi_r^2, \qquad t > 0, \quad \chi_r = \frac{1}{3} \left(r^2-1\right),
\end{cases}
\end{equation}
define an IBV problem for equation (\ref{kdv}) which is linearizable of type (b). 

In the case of linearizable boundary conditions, the formulation of the relevant RH problem requires only the analysis of the $x$-part of the Lax pair with the potential given in terms of the initial conditions.
In general, this analysis involves solving a linear Volterra integral equation. However, for the present example, due to the simple form of the initial profiles $E_r(x)$, the eigenfunctions of the $x$-part of the Lax pair can be expressed explicitly in terms of hypergeometric functions. This leads (via the formalism for linearizable IBV problems) to 
an explicit expression for the function $\Delta(k)$ (whose zeros correspond to the solitons present asymptotically in the solution $u_r(x,t)$). 
Even though the formula for $\Delta(k)$ is complicated, we can find the zeros numerically, and hence predict the number and shapes of solitons generated asymptotically by the initial and boundary data in (\ref{expIBVintro}). For example, we will find that for large $r$ there exists asymptotically only one stationary (time-independent) soliton, while as $r \downarrow 0$ the number of generated solitons appears to grow indefinitely. It is explained in section \ref{exponentialsec} that many of the relevant qualitative results can be explained in terms of the variations with respect to $r$ of the shape of the initial profile and of the boundary conditions. 

\subsubsection{Box-shaped intitial profiles}
Another natural and interesting class of initial profiles for which the $x$-part of the Lax pair can be solved explicitly is the class of box-shaped profiles defined as follows. Given data $I = \{h, L, x_0\} \subset \R \times \R^+ \times \R$, we let
\begin{equation*} 
\beta_I(x) = \begin{cases} 0, \qquad	0 < x < x_0, \\
h, \qquad	x_0 < x < x_0 + L,\\
0, \qquad  x_0 + L < x.
\end{cases}
\end{equation*}
The function $\beta_I(x)$ represents a box of height $h$ and length $L$ positioned with its left side at $x = x_0$.
The initial and boundary conditions
\begin{equation}\label{boxIBVintro}  
\begin{cases}
  u_I(x, 0) = \beta_I(x), \qquad x > 0, \\
  u_I(0,t) = u_{Ixx}(0,t) = 0, \qquad t > 0,
\end{cases}
\end{equation}
determine an IBV problem for equation (\ref{kdv}) which is linearizable of type (a). In this case we can again derive an explicit expression for the function $\Delta(k)$ and study numerically its set of zeros responsible for asymptotic solitons. 

Note that the functions $\beta_I(x)$ are well-defined also on the line and have decay as $x \to \pm \infty$ (this is in contrast to the case of the exponential initial profiles defined in (\ref{Erxdefintro}), which do not decay as $x \to -\infty$). Therefore, we may also apply the ordinary IST to determine the asymptotic behavior of the solution of (\ref{kdv}) with initial profile $\beta_I(x)$ in the absence of a boundary. Naturally, we find that the number and shapes of asymptotic solitons in the absence of a boundary is independent of the position $x_0$ of the box. However, in the presence of a boundary we find that the number of solitons {\it does depend} on $x_0$. The IBV problem with initial and boundary data given by (\ref{boxIBVintro}) therefore provides a concrete example elucidating the effect of the boundary on the asymptotics.

\subsection{Outline}
Before analyzing particular examples, we review in section \ref{spectralsec} the general theory of the IST for equation (\ref{kdv}) on the line and on the half-line. Our aim is to make the presentation as unified as possible, so that the similarities (and differences) of the line and half-line problems become apparent.

In section \ref{linearizablesec} we consider the special methods available for linearizable IBV problems. The steps required for this analysis were already outlined in \cite{Fbook}; we will state the final results but also consider certain aspects in greater detail.

In sections \ref{solitonsec} - \ref{boxsec} we consider the three main classes of examples: Solitons; IBV problems with exponential intitial profiles; and IBV problems with box-shaped initial profiles. 

In the appendix we give the formulas necessary relating the matrix and scalar Lax pairs associated with equation (\ref{kdv}).

\section{Spectral theory} \label{spectralsec} \nequation
In this section we review the spectral theory of equation (\ref{kdv}) on the line and on the half-line. A general introduction to the Inverse Scattering Transform can be found in \cite{A-C} for the case on the line, while the tools needed for the case on the half-line were first introduced in \cite{F1997, F2002} (see also \cite{Fbook}). Our treatment here seeks to implement the spectral theory for the line and half-line problems in a way that emphasizes the similarities between the two approaches. We will define four eigenfunctions $\{\mu_j\}_{1}^4$ of the Lax pair associated to (\ref{kdv}); the eigenfunctions $\mu_1, \mu_2, \mu_3$ will be used in the analysis of the half-line problem, whereas the infinite line problem utilizes the eigenfunctions $\mu_3$ and $\mu_4$. For both the infinite line and the half-line problem, we then express the solution $u(x,t)$ of equation (\ref{kdv}) in terms of the solution of a $2 \times 2$ matrix Riemann-Hilbert problem.

\subsection{Lax pair}
Let
$$\sigma_1 = \begin{pmatrix} 0	&	1 \\ 1	& 0\end{pmatrix}, \qquad
\sigma_2 = \begin{pmatrix} 0	&	-i \\ i	& 0\end{pmatrix}, \qquad
\sigma_3 = \begin{pmatrix} 1	&	0 \\ 0	& -1\end{pmatrix}.$$
Equation (\ref{kdv}) admits the Lax pair formulation
\begin{equation}\label{lax}
\begin{cases}	& \mu_x - ik[\sigma_3, \mu] = V_1\mu,
			\\
	& \mu_t + i(k + 4k^3)[\sigma_3, \mu] = V_2\mu,
\end{cases}
\end{equation}
where $k \in \C$ is a spectral parameter, $\mu(x,t,k)$ is a $2 \times 2$ matrix valued eigenfunction, and
\begin{align}\label{V1V2def}
&V_1(x,t,k) = \frac{u}{2k}(\sigma_2 - i\sigma_3),
	\\
&V_2(x,t,k) = -2ku\sigma_2 + u_x \sigma_1 + \frac{2u^2 + u - u_{xx}}{2k} (i\sigma_3 - \sigma_2).
\end{align}
We can write the Lax pair (\ref{lax}) in differential form as
\begin{equation}\label{laxdiffform}  
  d\left(e^{i(-k x + (k + 4k^3) t)\hat{\sigma}_3} \mu(x,t,k)\right) = W(x,t,k),
\end{equation}
where
$$W = e^{i(-kx + (k + 4k^3)t)\hat{\sigma}_3}\left(V_1\mu dx + V_2 \mu dt\right),$$
and $\hat{\sigma}_3$ acts on a $2\times 2$ matrix $A$ by $\hat{\sigma}_3A = [\sigma_3, A], e^{\hat{\sigma}_3} A = e^{\sigma_3} A e^{-\sigma_3}$.

\subsection{Bounded and analytic eigenfunctions}\label{boundedanalyticsubsec}
\begin{figure}
\begin{center}
    \includegraphics[width=.4\textwidth]{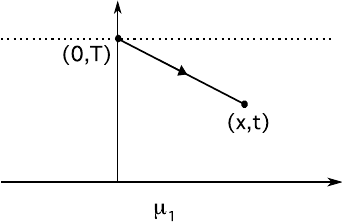} \quad
   \includegraphics[width=.4\textwidth]{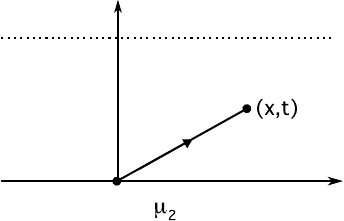} \quad
   \includegraphics[width=.4\textwidth]{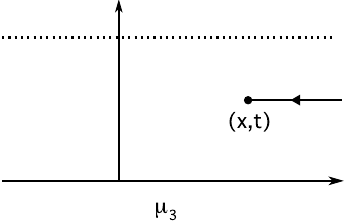} \quad
    \includegraphics[width=.4\textwidth]{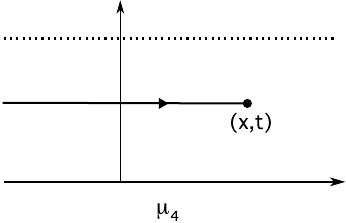} \\
     \begin{figuretext}\label{mu1234.pdf}
       The contours of integration for the solutions $\mu_1$, $\mu_2$, $\mu_3$, and $\mu_4$ of (\ref{lax}).
     \end{figuretext}
     \end{center}
\end{figure}

Let equation (\ref{laxdiffform}) be valid for $0 < t< T$ and $-\infty < x < \infty$ ($0 < x < \infty$ when considering the half-line problem)
where $T \leq \infty$ is a given positive constant. Unless otherwise specified, we suppose that the solutions are defined for $-\infty < x < \infty$; we will indicate what changes are necessary in the case of the half-line problem.
Assuming that the function $u(x,t)$ has sufficient smoothness and decay, we introduce four solutions $\mu_j$, $j = 1,2,3,4$, of (\ref{laxdiffform}) by
\begin{equation}\label{mujdef}  
  \mu_j(x,t,k) = I + \int_{(x_j, t_j)}^{(x,t)} e^{-i(-kx + (k+4k^3)t)\hat{\sigma}_3}W(x',t',k),
\end{equation}
where $(x_1, t_1) = (0, T)$, $(x_2, t_2) = (0, 0)$, $(x_3, t_3) = (\infty, t)$, and $(x_4, t_4) = (-\infty, t)$. 
The function $\mu_4$ is only defined when considering the infinite line problem. If $T = \infty$, the function $\mu_1$ is only defined if $u(0,t)$ decays to zero as $t \to \infty$. Since the one-form $W$ is exact, the integral on the right-hand side of (\ref{mujdef}) is independent of the path of integration. We choose the particular contours shown in Figure \ref{mu1234.pdf}. This choice implies the following inequalities on the contours,
\begin{align*}
(x_1, t_1) \to (x, t): x' - x \leq 0,& \qquad t' - t \geq 0,
	\\
(x_2, t_2) \to (x, t): x' - x \leq 0,& \qquad t' - t \leq 0,
	\\
(x_3, t_3) \to (x, t): x' - x \geq 0,&
	\\
(x_4, t_4) \to (x, t): x' - x \leq 0.&
\end{align*}
The second column of the matrix equation (\ref{mujdef}) involves $\exp[2i(-k(x' - x) + (k + 4k^3)(t' - t))]$. 
Using the above inequalities it follows that this exponential is bounded in the following regions of the complex $k$-plane,
\begin{align*}
(x_1, t_1) \to (x, t): \{\text{Im}\, k \geq 0\} &\cap \{\text{Im}\, [k + 4k^3] \geq 0\},
	\\
(x_2, t_2) \to (x, t): \{\text{Im}\, k \geq 0\} &\cap \{\text{Im}\, [k + 4k^3] \leq 0\},
	\\
(x_3, t_3) \to (x, t): \{\text{Im}\, k\leq 0\}&,
	\\
(x_4, t_4) \to (x, t): \{\text{Im}\, k \geq 0\}&.
\end{align*}
We define the sets (see Figure \ref{D1234.pdf})
\begin{align*}
&D_1 = \{k \in \C| \text{Im}\, k < 0  \text{  and  } \text{Im}\, [k + 4k^3] > 0 \},
	\\
&D_2 = \{k \in \C| \text{Im}\, k < 0  \text{  and  } \text{Im}\, [k + 4k^3] < 0\},
	\\
&D_3 = \{k \in \C|\text{Im}\, k > 0  \text{  and  } \text{Im}\, [k + 4k^3] > 0\},
	\\
&D_4 = \{k \in \C| \text{Im}\, k > 0  \text{  and  } \text{Im}\, [k + 4k^3] < 0\}.
\end{align*}
The second column vectors of $\mu_1$, $\mu_2$, $\mu_3$, $\mu_4$ are bounded and analytic for $k \in \C$ such that $k$ belongs to $D_3$, $D_4$, $D_1 \cup D_2$, and $D_3 \cup D_4$, respectively.
We will denote these vectors with superscripts $(3)$, $(4)$, $(12)$, and $(34)$ to indicate these boundedness properties.
Similar conditions are valid for the first column vectors. We obtain
$$\mu_1 = \left(\mu_1^{(2)}, \mu_1^{(3)}\right), \quad 
\mu_2 = \left(\mu_2^{(1)}, \mu_2^{(4)}\right), \quad 
\mu_3 = \left(\mu_3^{(34)}, \mu_3^{(12)}\right), \quad 
\mu_4 = \left(\mu_4^{(12)}, \mu_4^{(34)}\right)
.$$
We deduce from (\ref{mujdef}) that $\mu_1$ is an entire function of $k$, and that $\mu_2$ is an entire function of $k$ if $T < \infty$. 

Let us collect some properties of the eigenfunctions of (\ref{lax}). 
As $k \to 0$, a solution $\mu(x,t,k)$ of (\ref{lax}) satisfies 
\begin{equation}\label{muatorigin}  
  \mu(x,t,k) = \frac{i \alpha(x,t)}{k}\begin{pmatrix} 1 & 1 \\ -1 & -1 \end{pmatrix} + O(1), \quad k \to 0,
\end{equation}  
where $\alpha(x,t)$ is a real-valued function cf. \cite{Fbook}.
From the form of the matrices $V_1$ and $V_2$ defined in (\ref{V1V2def}), it follows that each eigenfunction $\mu$ of (\ref{lax}) obeys the following symmetries:
\begin{align}\label{musymmetries}
& \mu_{11}(x,t,k) = \overline{\mu_{22}(x,t,\bar{k})}, \qquad \mu_{12}(x,t,k) = \overline{\mu_{21}(x,t,\bar{k})},
	\\
& \mu_{11}(x,t,k) = \mu_{22}(x,t,-k), \qquad \mu_{12}(x,t,k) = \mu_{21}(x,t,-k).
\end{align}
It also holds that
$$\mu_j(x, t, k) = I + O\left(\frac{1}{k}\right), \qquad k \to \infty, \quad j = 1,2,3,4.$$
Furthermore, note that
\begin{equation}\label{detmuisone}  
  \det \mu_j = 1, \qquad j= 1,2,3,4,
\end{equation}
and that any two solutions $\mu$ and $\tilde{\mu}$ of (\ref{laxdiffform}) are related by an equation of the form
\begin{equation}\label{mutildemu}  
  \mu(x,t,k) = \tilde{\mu}(x,t,k)  e^{-i(-kx + (k+4k^3)t)\hat{\sigma}_3} C_0(k),
\end{equation}
where $C_0(k)$ is a $2\times 2$ matrix independent of $x$ and $t$.

The $\mu_j$'s are the fundamental eigenfunctions needed for the formulation of a Riemann-Hilbert problem in the complex $k$-plane. The eigenfunctions $\mu_1, \mu_2, \mu_3$ are used for the half-line problem, while $\mu_3, \mu_4$ are used for the infinite line problem.

\subsection{Spectral theory on the line}
We now restrict attention to the infinite line problem. In view of (\ref{mutildemu}), we may define a spectral function $s^L(k)$ by\footnote{Some quantities related to the problem on the line will be denoted with the superscript $L$ to distinguish them from their half-line counterparts.} 
\begin{equation}\label{lineseq} 
  \mu_3(x,t,k) = \mu_4(x,t,k)e^{-i(-k x + (k+4k^3)t) \hat{\sigma}_3} s^{L}(k), \qquad \text{Im}\, k = 0.
\end{equation}
Evaluation of (\ref{lineseq}) at $t = 0$ and $x \to -\infty$ gives
\begin{equation}\label{linesexplicit}
  s^{L}(k) = I - \int_{-\infty}^\infty e^{-i kx \hat{\sigma}_3} (V_1 \mu_3)(x,0, k)dx, \qquad \text{Im}\, k = 0.
\end{equation}
Moreover, by (\ref{detmuisone}),
$$\det s^{L}(k) = 1.$$
We infer from the symmetries (\ref{musymmetries}) that there exist functions $a^{L}(k)$ and $b^{L}(k)$ such that
\begin{equation}\label{linesabmatrix}
s^{L}(k) = \begin{pmatrix} 
\overline{a^{L}(\bar{k})} 	&	b^{L}(k)	\\
\overline{b^{L}(\bar{k})}	&	a^{L}(k)
\end{pmatrix}.
\end{equation}
From the explicit expression (\ref{linesexplicit}) for $s^{L}(k)$ and the fact that the second column of $\mu_3$ is defined and analytic in $D_1 \cup D_2$, we deduce that $a^{L}(k)$ has an analytic continuation to $D_1 \cup D_2$. It follows from (\ref{muatorigin}) and (\ref{lineseq}) that 
$$a^L(k) = \frac{i \alpha}{k} + O(1), \quad b^L(k) = -\frac{i\alpha}{k} + O(1), \qquad k \to 0,$$
for some real constant $\alpha$.

\subsubsection{Residue conditions}
We assume that $a^{L}(k)$ has $N$ simple zeros $\{k_j\}_{j = 1}^{N}$ in the lower half-plane. These zeros automatically lie on the imaginary axis as we consider only real-valued solutions of (\ref{kdv}). This can be seen as a consequence of the fact that the operator defining the $x$-part of the scalar Lax pair in (\ref{scalarlax}) with eigenvalue $k^2$ is self-adjoint and so has a purely real spectrum.

The second column of equation (\ref{lineseq}) is
\begin{equation}\label{linemu2amu1b}
  \mu_3^{(12)} = a^{L}\mu_4^{(34)} + b^{L}\mu_4^{(12)}e^{2ik x}, \qquad  \text{Im}\, k = 0.
\end{equation}
Applying $\det \left(\mu_4^{(12)}, \cdot\right)$ to this equation and recalling (\ref{detmuisone}), we find
$$\det\left(\mu_4^{(12)}(x, t, k),  \mu_3^{(12)}(x, t, k)\right) = a^{L}(k), \qquad k \in \bar{D}_1 \cup \bar{D}_2,$$
where we have used that both sides are well-defined and analytic in the lower half-plane to extend the above relation to $\bar{D}_1 \cup \bar{D}_2$.
Hence, if $a^{L}(k_j) = 0$, then $\mu_4^{(12)}(x, t, k_j)$ and $\mu_3^{(12)} (x, t, k_j)$ are linearly dependent vectors for each $x$ and $t$. It follows that there exist constants $b_j$ such that
\begin{equation}\label{linemubj}  
  \mu_4^{(12)}(x, t, k_j) = b_j e^{2i (-k_j x + (k_j +4k_j^3)t)} \mu_3^{(12)}(x, t, k_j), \qquad x \in \R, \, t > 0.
\end{equation}
Recalling the symmetries (\ref{musymmetries}), the complex conjugate of (\ref{linemubj}) is
$$\mu_4^{(34)}(x, t, \bar{k}_j) = \bar{b}_j e^{-2i (-\bar{k}_j x + (\bar{k}_j +4\bar{k}_j^3)t)} \mu_3^{(34)}(x, t, \bar{k}_j), \qquad x \in \R, \, t > 0.$$
Consequently, the residues of $\mu_4^{(12)}/a^{L}$ and $\mu_4^{(34)}/\bar{a}^{L}$ at $k_j$ and $\bar{k}_j$ are
\begin{align*}
& \underset{k_j}{\text{Res}} \frac{\mu_4^{(12)}(x,t, k)}{a^L(k)} = \frac{\mu_4^{(12)}(x, t, k_j)}{\dot{a}^{L}(k_j)} =  C_j e^{2i (-k_j x + (k_j +4k_j^3)t)} \mu_3^{(12)}(x, t, k_j),
	\\
& \underset{\bar{k}_j}{\text{Res}}  \frac{\mu_4^{(34)}(x,t,k)}{\overline{a^L(\bar{k})}} = \frac{\mu_4^{(34)}(x, t, \bar{k}_j)}{\overline{\dot{a}^{L}(k_j)}} =  \bar{C}_j e^{-2i (-\bar{k}_j x + (\bar{k}_j +4\bar{k}_j^3)t)} \mu_3^{(34)}(x, t, \bar{k}_j),
\end{align*}
where $\dot{a}^{L} = \frac{da^{L}}{dk}$ and $C_j = \frac{b_j}{\dot{a}^{L}(k_j)}$.

\begin{remark} \upshape
In the traditional approach to the Inverse Scattering Transform the $x$-part of the Lax pair (\ref{lax}) is first analyzed at each fixed time $t$ to yield time-dependent scattering data consisting of two functions $a^{L}(k,t)$ and $b^{L}(k,t)$, together with the zeros $\{k_j(t)\}_1^N$ of $a^{L}(k,t)$ and the corresponding normalization constants $\{C_j(t)\}_1^N$. The $t$-part of the Lax pair is then used to determine the time-evolution of the scattering data. In our present formulation the $x$- and $t$-parts of the Lax pair are analyzed simultaneously, which implies that no separate analysis is necessary in order to find the time-evolution. The relation between the scattering data in the traditional analysis $\{a^{L}(k, t), b^{L}(k, t), \{k_j(t)\}_1^N, \{C_j(t)\}_1^N\}$ and the data $\{a^{L}(k), b^{L}(k), \{k_j\}_1^N, \{C_j\}_1^N\}$ in our current treatment is 
\begin{align*}
&a^{L}(k,t) = a^{L}(k), \qquad b^{L}(k,t) = b^{L}(k)e^{-2i(k+4k^3)t}, 
	\\
&k_j(t) = k_j, \qquad C_j(t) = C_j e^{2i(k_j + 4k_j^3) t}.
\end{align*}
In our current treatment the time-dependent exponential factors $e^{\pm 2i(k_j + 4k_j^3) t}$ are automatically built into the equations (\ref{lineseq}) and (\ref{linemubj}).
We have chosen the present approach for two reasons: (i) It is the proper approach for dealing with initial-boundary value problems, so that adopting it also for the problem on the line makes the link to the half-line problem more visible. (ii) We believe it is more natural also for the problem on the line.
\end{remark}

\subsubsection{Riemann-Hilbert problem}
We now describe how the solution to the initial-value problem for equation (\ref{kdv}) on the line can be expressed in terms of the solution $M(x,t,k)$ of a $2 \times 2$ matrix Riemann-Hilbert problem.

By substituting the expansion
$$\mu = I + \frac{m^{(1)}}{k} + \frac{m^{(2)}}{k^2} + O\left(\frac{1}{k^3}\right), \qquad k \to \infty,$$
in the $x$-part of (\ref{lax}), we find by considering the terms of $O(1/k)$ that
$$u(x) = -2im^{(1)}_{22x}.$$
Algebraic manipulations show that relation (\ref{lineseq}) can be rewritten in the form of the RH problem
$$M_-(x,t,k) = M_+(x, t, k)J(x, t, k), \qquad \text{Im}\, k = 0,$$
where the matrices $M_-$, $M_+$, $J$ are defined by
\begin{align}\label{MplusMminus}
M_+ =& \left(\frac{\mu_4^{(12)}}{a^{L}(k)}, \mu_3^{(12)}\right), \qquad \text{Im}\, k \leq 0;
	\\ \nonumber
M_- =& \left(\mu_3^{(34)}, \frac{\mu_4^{(34)}}{\overline{a^{L}(\bar{k})}}\right), \qquad \text{Im}\, k \geq 0;
	\\
\label{lineJdef}
J =& \begin{pmatrix} 1	&	-\frac{b^{L}(k)}{\overline{a^{L}(\bar{k})}}e^{-2i(-kx + (k + 4k^3) t)} 	\\
\frac{\overline{b^{L}(\bar{k})}}{a^{L}(k)} e^{2i(-kx + (k + 4k^3) t)}  	&	\frac{1}{a^{L}(k)\overline{a^{L}(\bar{k})}} \end{pmatrix}, \qquad \text{Im}\, k = 0.
\end{align}
The contour for this RH problem is the real axis. 

We summarize our discussion of spectral theory on the line in the following theorem, which reduces the Cauchy problem  for equation (\ref{kdv}) to a $2 \times 2$ matrix Riemann-Hilbert problem.
We use the notation $[A]_1$ ($[A]_2$) for the first (second) column of a $2\times 2$ matrix $A$. 

\begin{theorem}\label{lineRHtheorem}
Given initial data $u_0(x)$ such that $u_0(x), xu_0(x) \in L^1(\R^+)$, let $V_1 = \frac{u_0}{2k}(\sigma_2 - i\sigma_3)$ and let
$\mu_3(x,0, k)$ and $\mu_4(x,0, k)$ be the unique solutions of the following Volterra linear integral equations:
\begin{align*} 
  \mu_3(x,0,k) = I - \int_x^{\infty} e^{-i k (x' - x) \hat{\sigma}_3} (V_1 \mu_3)(x',0,k)dx',
 	\\ 
  \mu_4(x,0,k) = I + \int_{-\infty}^x e^{-i k(x' - x) \hat{\sigma}_3} (V_1 \mu_4)(x',0,k)dx'.
\end{align*}
Define $\{a^{L}(k), b^{L}(k), C_j\}$ by
\begin{equation}\label{abRH}
  \begin{pmatrix} b^{L}(k) \\
 a^{L}(k) \end{pmatrix} = [s^{L}(k)]_2, \qquad s^{L}(k) = I - \int_{-\infty}^\infty e^{-i kx \hat{\sigma}_3} (V_1 \mu_3)(x,0, k)dx, 		\qquad \text{\upshape Im}\, k = 0,
\end{equation}
and
\begin{equation}\label{CjRH}
  [\mu_4(x, 0, k_j)]_1 = \dot{a}^{L}(k_j) C_j e^{- 2i k_j x} [\mu_3(x, 0, k_j)]_2,  \qquad j = 1, \dots, N,
 \end{equation}
where we assume that $a^{L}(k)$ has $N$ simple zeros $\{k_j\}_{j = 1}^{N} \subset i \R^-$. 

Then
\begin{itemize}
\item $a^{L}(k)$ is defined for $\text{Im}\, k \leq 0$ and analytic in $\text{Im}\, k < 0$.
\item $b^{L}(k)$ is defined for $k \in \R$.
\item $a^{L}(k)\overline{a^{L}(\bar{k})} - b^{L}(k)\overline{b^{L}(\bar{k})} = 1, \qquad k \in \R$.
\item $a^{L}(k) = \overline{a^L(-\bar{k})}, \qquad \text{Im}\, k \leq 0$.
\end{itemize}
Moreover, the solution $u(x,t)$ of equation (\ref{kdv}) with initial data $u_0(x)$ is given by
\begin{equation}\label{linerecoveru}
  u(x, t) = -2 i \lim_{k \to \infty} k \partial_x\left( M_{22}(x,t,k)\right),
\end{equation}
where $M(x, t, k)$ is the unique solution of the following RH problem:
\begin{itemize}
\item $M(x,t,k) = \left\{ \begin{array}{ll}
M_+(x,t,k), \qquad \text{Im}\, k \leq 0, \\
M_-(x,t,k), \qquad \text{Im}\, k \geq 0, \\
\end{array} \right.$ 

is a sectionally meromorphic function.

\item $M_-(x,t,k) = M_+(x, t, k)J(x,t, k)$ for $k \in \R,$ where $J$ is defined in (\ref{lineJdef}).

\item $M(x, t, k)$ has the asymptotic behavior
\begin{equation}\label{MtoI}
M(x, t, k) = I + O\left(\frac{1}{k}\right), \qquad k \to \infty.
\end{equation}

\item For some real function $\alpha(x, t)$, 
\begin{align}\label{lineMplusatorigin}
  &M_+(x, t, k) \sim \frac{i \alpha(x, t)}{k}\begin{pmatrix} 0 & 1 \\ 0 & -1 \end{pmatrix}, \qquad k \to 0
  	\\ \label{lineMminusatorigin}
  &M_-(x, t, k) \sim \frac{i \alpha(x, t)}{k}\begin{pmatrix} 1 & 0 \\ -1 & 0 \end{pmatrix}, \qquad k \to 0.
\end{align}

\item For some real constant $\alpha$,
$$a^L(k) = \frac{i \alpha}{k} + O(1), \quad b^L(k) = -\frac{i\alpha}{k} + O(1), \qquad k \to 0.$$

\item The first column of $M_+$ has simple poles at $k = k_j$, $j = 1, \dots, N$, and the second column of $M_-$ has simple poles at $k = \bar{k}_j$, $j = 1, \dots, N$.
The associated residues are given by
\begin{align}\label{lineresidue1}
\underset{k_j}{\text{\upshape Res}} [M(x,t,k)]_1 =& C_j e^{2i (-k_j x + (k_j +4k_j^3)t)} [M(x, t, k_j)]_2, \qquad j = 1, \dots, N,
		\\\label{lineresidue2}
\underset{\bar{k}_j}{\text{\upshape Res}} [M(x,t,k)]_2 =& \bar{C}_j e^ {-2i (-\bar{k}_j x + (\bar{k}_j +4\bar{k}_j^3)t)} [M(x, t, \bar{k}_j)]_1, \qquad j = 1, \dots, N.
\end{align}

\end{itemize}

\end{theorem}

\begin{remark}\label{regularizedremark}{\bf (Regularized RH problem)} \upshape
The RH problem of Theorem \ref{lineRHtheorem} has a singularity at $k = 0$ (see equations (\ref{lineMplusatorigin}) and (\ref{lineMminusatorigin})). It is sometimes more convenient to consider the regularized RH problem with no singularity at $k=0$. The solutions of the two problems are related as follows: If $\check{M}(x, t, k)$ is a solution of the regularized RH problem (i.e. $\check{M}$ satisfies the jump and residue conditions, but is regular at $k = 0$), then 
\begin{equation}\label{MfromregularM}
M(x, t, k) = \left(I + \frac{iy(x, t)}{2k}\begin{pmatrix} 1 & 1 \\ -1 & -1 \end{pmatrix}\right)\check{M}(x, t, k), \qquad y(x, t) = 2i\lim_{k \to \infty} (k\check{M}_{12}(x,t, k)),
\end{equation}
is a solution of the RH problem of Theorem \ref{lineRHtheorem} cf. \cite{FI94}.
\end{remark}

\begin{remark} {\bf (Symmetries of the solution of the RH problem))} \upshape
It follows from the symmetries (\ref{musymmetries}) that the solution $M(x, t,  k)$ of the Riemann-Hilbert problem in Theorem \ref{lineRHtheorem} respects the symmetries
\begin{align}\label{Msymmetry}
& M_{11}(x, t, k) = \overline{M_{22}(x,  t, \bar{k})}, \qquad M_{21}(x, t, k) = \overline{M_{12}(x,  t, \bar{k})},
	\\ \nonumber
 & M_{11}(x, t, k) = M_{22}(x, t, -k), \qquad M_{12}(x, t,  k) = M_{21}(x, t, -k).
\end{align}
Moreover, if one enforces (\ref{Msymmetry}), then only one of the two residue conditions (\ref{residue1})-(\ref{residue2}) needs to be verified since the other condition is a consequence of symmetry.
\end{remark}

\subsubsection{Soliton solutions}\label{solitonsubsec} 
We next outline how to find the solitons of equation (\ref{kdv}) and the corresponding eigenfunctions of the Lax pair (\ref{lax}). We will follow these steps in subsections \ref{onesolitonsubsec} and \ref{twosolitonsubsec} below to compute the one and two-solitons and their associated eigenfunctions.

The solitons correspond to spectral data $\{a^L(k), b^L(k), C_j\}$ for which $b^L(k)$ vanishes identically. In this case the jump matrix $J$ in (\ref{lineJdef}) is the identity matrix and the RH problem of Theorem \ref{lineRHtheorem} consists of finding a meromorphic function $M(x,t,k)$ satisfying (\ref{MtoI}), the residue conditions (\ref{lineresidue1})-(\ref{lineresidue2}), and the conditions (\ref{lineMplusatorigin})-(\ref{lineMminusatorigin}) at $k = 0$. We first construct the solution $\check{M}$ to the regularized RH problem with no singularity at $k=0$; subsequently, the solution $M$ is found from (\ref{MfromregularM}). In what follows we will suppress the $(x,t)$-dependence of $\check{M}$ and write $\check{M}(k)$ for $\check{M}(x,t,k)$.

Let $\theta_j = -k_j x + (k_j +4k_j^3)t$. From (\ref{MtoI}) and (\ref{lineresidue1}) we find
\begin{equation}\label{M1decompose}
[\check{M}(k)]_1 = \begin{pmatrix} 1	\\	0 \end{pmatrix} + \sum_{j = 1}^{N} \frac{C_j e^{2i\theta_j} [\check{M}(k_j)]_2}{k - k_j}.
\end{equation}
If we impose the symmetries (\ref{Msymmetry}), equation (\ref{M1decompose}) can be written as
\begin{equation}\label{M2kfromM2kj}
\begin{pmatrix} \overline{\check{M}_{22}(\bar{k})} \\ \overline{\check{M}_{12}( \bar{k})} \end{pmatrix} = \begin{pmatrix} 1	\\	0 \end{pmatrix} + \sum_{j = 1}^{N} \frac{C_j e^{2i\theta_j}}{k - k_j}\begin{pmatrix} \check{M}_{12}(k_j) \\ \check{M}_{22}(k_j) \end{pmatrix}.
\end{equation}
Evaluation at $\bar{k}_n$ yields 
\begin{equation}\label{algebraicsystem}
\begin{pmatrix} \overline{\check{M}_{22}(k_n)} \\ \overline{\check{M}_{12}( k_n)} \end{pmatrix} = \begin{pmatrix} 1	\\	0 \end{pmatrix} + \sum_{j = 1}^{N} \frac{C_j e^{2i\theta_j}}{\bar{k}_n - k_j}\begin{pmatrix} \check{M}_{12}( k_j) \\ \check{M}_{22}(k_j) \end{pmatrix}, \qquad n =1, \dots, N.
\end{equation}
Solving this algebraic system for $\check{M}_{12}(k_j)$ and $\check{M}_{22}(k_j)$, $j = 1, \dots, N$, and substituting the result back into (\ref{M2kfromM2kj}) gives an explicit expression for $[\check{M}(k)]_2$. The first column of $\check{M}(k)$ is obtained by symmetry. This construction provides the solution of the regularized Riemann-Hilbert problem. The solution $M$ of the original RH problem is found from (\ref{MfromregularM}), and the soliton solution $u(x,t)$ is obtained from (\ref{linerecoveru}).

\subsection{Spectral theory on the half-line}\label{spectralhalflinesubsec}
We now turn to the spectral analysis for the half-line problem. The presentation will be brief; see \cite{Fbook} for further details. 
We define $s(k)$ and $S(k)$ by the relations
\begin{align}
\label{seq} 
  \mu_3(x,t,k) &= \mu_2(x,t,k)e^{-i (-k x + (k +4k^3)t)\hat{\sigma}_3} s(k),
		\\
  \label{Seq} 
  \mu_1(x,t,k) &= \mu_2(x,t,k)  e^{-i (-k x + (k +4k^3)t)\hat{\sigma}_3} S(k).
\end{align}
Evaluation of (\ref{seq}) and (\ref{Seq}) at $(x,t) = (0,0)$ and $(x,t) = (0,T)$ gives the expressions
\begin{equation}\label{Ssdef}   
   s(k) = \mu_3(0,0, k), \qquad S(k) = \mu_1(0,0, k) = \left(e^{i(k + 4k^3)T\hat{\sigma}_3}\mu_2(0,T,k)\right)^{-1},
\end{equation}
where the final equality is valid only if $T < \infty$. We use the following notation for $s$ and $S$:
\begin{equation}\label{sSandabAB}
s(k) = \begin{pmatrix} \overline{a(\bar{k})} & b(k) \\
 \overline{b(\bar{k})} 	&	a(k) \end{pmatrix}, \qquad 
S(k) = \begin{pmatrix} \overline{A(\bar{k})} & B(k) \\
 \overline{B(\bar{k})} 	&	A(k) \end{pmatrix}.
 \end{equation}
The spectral functions $a(k)$ and $b(k)$ have the following properties:
\begin{enumerate}
\item[(i)] $a(k)$ and $b(k)$ are continuous and bounded for $k \in \bar{D}_1 \cup \bar{D}_2$ and analytic in the interior of this set.
\item[(ii)] $a(k) = 1 + O(1/k), \quad b(k) = O(1/k), \qquad k \to \infty, \quad k \in D_1 \cup D_2$.
\item[(iii)] $|a(k)|^2 - |b(k)|^2 = 1, \qquad k \in \bar{D}_1 \cup \bar{D}_2$.
\item[(iv)] For some real constant $\alpha$,
\begin{equation}\label{abkto0}
a(k) = \frac{i \alpha}{k} + O(1), \quad b(k) = -\frac{i\alpha}{k} + O(1), \qquad k \to 0.
\end{equation}
\end{enumerate}
The spectral functions $A(k)$ and $B(k)$ have the following properties:
\begin{enumerate}
\item[(i)] $A(k)$ and $B(k)$ are entire functions bounded for $k \in \bar{D}_1 \cup \bar{D}_3$ when $T < \infty$. If $T = \infty$, the functions $A(k)$ and $B(k)$ are defined only for $k \in \bar{D}_1 \cup \bar{D}_3$.
\item[(ii)] $A(k) = 1 + O(1/k), \quad B(k) = O(1/k), \qquad k \to \infty, \quad k \in D_1 \cup D_3$.
\item[(iii)] $A(k)\overline{A(\bar{k})} -  B(k)\overline{B(\bar{k})} = 1, \qquad k \in \begin{cases}
\C, & T < \infty, \\  \bar{D}_1 \cup \bar{D}_3, & T = \infty.\end{cases}$
\item[(iv)] For some real constant $\beta$,
\begin{equation}\label{ABkto0}
A(k) = \frac{i \beta}{k} + O(1), \quad B(k) = -\frac{i \beta}{k} + O(1), \qquad k \to 0.\end{equation}
\end{enumerate}

\subsubsection{Global relation}
Applying Stokes' theorem to the domain $\{0 < x < \infty, 0 < t < T\}$ and the closed one-form $W$ with $\mu = \mu_3$, one finds the following so-called global relation (see \cite{Fbook}):
\begin{equation}\label{globalrelation}
   B(k)a(k) - A(k)b(k) = \begin{cases} e^{2i(k + 4k^3) T} c^+(k) \quad \text{for} \quad k \in \bar{D}_1 \cup \bar{D}_2,& T < \infty, \\
       0\quad \text{for} \quad  k \in \bar{D}_1,& T = \infty,
       \end{cases}
\end{equation}
where
$$c^+(k) = \int_0^\infty e^{-2ik x' }  (V_1\mu_3)_{12}(x', T, k)dx'.$$

\subsubsection{Riemann-Hilbert problem}
Equations (\ref{seq}) and (\ref{Seq}) can be rewritten in the following form, expressing the jump condition of a $2 \times 2$ RH problem:
$$M_-(x,t,k) = M_+(x,t, k)J(x,t,k), \qquad k \in \bar{D}_i \cap \bar{D}_j, \quad i,j = 1, \dots, 4,$$
where the matrices $M_-$, $M_+$, and $J$ are defined by
\begin{align}\label{MplusMminusdef}
M_+ = \left(\frac{\mu_2^{(1)}}{a(k)}, \mu_3^{(12)}\right), \quad k \in \bar{D}_1; \qquad
M_- = \left(\frac{\mu_1^{(2)}}{d(k)}, \mu_3^{(12)}\right), \quad k \in \bar{D}_2;
		\\ \nonumber
M_+ = \left(\mu_3^{(34)}, \frac{\mu_1^{(3)}}{\overline{d(\bar{k})}}\right), \quad k \in \bar{D}_3; \qquad
M_- = \left(\mu_3^{(34)}, \frac{\mu_2^{(4)}}{\overline{a(\bar{k})}}\right), \quad k \in \bar{D}_4;
\end{align}
\begin{equation}\label{ddef}
  d(k) =a(k)\overline{A(\bar{k})} -  b(k) \overline{B(\bar{k})}, \qquad k \in \bar{D}_2;
\end{equation}
\begin{equation}
J(x,t,k) = \left\{ \begin{array}{ll}
J_1 & k \in \bar{D}_1 \cap \bar{D}_2  \\
J_2 = J_3 J_4^{-1} J_1 & k \in \bar{D}_2 \cap \bar{D}_3 \\
J_3 & k \in \bar{D}_3 \cap \bar{D}_4 \\
J_4 & k \in \bar{D}_4 \cap \bar{D}_1 \\
\end{array} \right.
\end{equation}
\begin{figure}
\begin{center}
   \includegraphics[width=.5\textwidth]{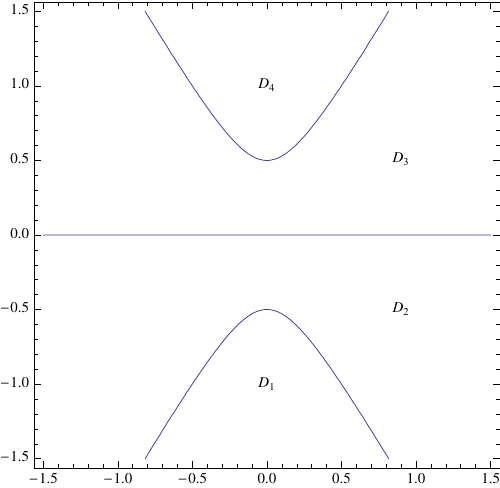} \\
     \begin{figuretext}\label{D1234.pdf}
        The contour for the Riemann-Hilbert problem in the complex $k$-plane.
     \end{figuretext}
     \end{center}
\end{figure}
with
\begin{equation}\label{J123def}
J_1 = \begin{pmatrix} 1	&	0 	\\
\Gamma(k)e^{2i\theta(k)}	&	1 \end{pmatrix}, \quad 
J_4 = \begin{pmatrix} 1	&	-\frac{b(k)}{\overline{a(\bar{k})}}e^{-2i\theta(k)} 	\\
 \frac{\overline{b(\bar{k})}}{a(k)} e^{2i\theta(k)}	&	\frac{1}{a(k)\overline{a(\bar{k})}} \end{pmatrix}, \quad 
J_3 = \begin{pmatrix} 1	&	- \overline{\Gamma(\bar{k})}e^{-2i\theta(k)} 	\\
0	&	1 \end{pmatrix};
\end{equation}
\begin{equation}\label{Gammadef}
\theta(x,t,k) = -k x + (k +4k^3)t; \qquad \Gamma(k) = \frac{\overline{B(\bar{k})}}{a(k)d(k)}, \quad k \in \bar{D}_2.
\end{equation}

The contour for this RH problem is depicted in Figure \ref{D1234.pdf}.

\begin{assumption}\label{zerosassumption}\upshape We assume that
\item[(i)] $a(k)$ has $N$ simple zeros $\{k_j\}_{j = 1}^{N} \subset i \R^-$ such that $k_j \in D_1$, $j = 1, \dots, n_1$, and $k_j \in D_2$, $j = n_1+1, \dots, N$. 

\item[(ii)] $d(k)$ has $\Lambda$ simple zeros $\{\lambda_j\}_1^{\Lambda}\subset i \R^-$, such that 
$\lambda_j \in D_2$, $j = 1, \dots, \Lambda.$ 

\item[(iii)] None of the zeros of $a(k)$ coincides with a zero of $d(k)$.
\end{assumption}

\begin{theorem}\label{RHtheorem}
Let $u_0(x)$, $x \geq 0$, and $\{g_j(t)\}_1^3$, $t \in [0, T)$, be such that $u_0(x), xu_0(x) \in L^1(\R^+)$ and $g_j(t), tg_j(t) \in L^1([0, T))$, $j = 1,2,3$. 
Define the spectral functions $a(k)$, $b(k)$, $A(k)$, and $B(k)$ according to (\ref{Ssdef}) and (\ref{sSandabAB}), where $\mu_1(0,t, k)$ and $\mu_3(x,0, k)$ are obtained as the unique solutions of the Volterra linear integral equations 
\begin{align} \label{mu1Volterra}
   \mu_1(0,t,k) = I + \int_{T}^{t} e^{i(k+4k^3) (t' -t)\hat{\sigma}_3}  (V_2 \mu_1)(0,t',k)dt'.
	\\ \label{mu3Volterra}
  \mu_3(x,0,k) = I - \int_x^{\infty} e^{-i k (x' - x) \hat{\sigma}_3} (V_1 \mu_3)(x',0,k)dx',
\end{align}
and $V_1(x, 0,\zeta)$, $V_2(0,t,\zeta)$ are given by equation (\ref{V1V2def}) in terms of the initial and boundary values
$$u_0(x) = u(x,0), \qquad  g_0(t) = u(0, t), \qquad  g_1(t) = u_x(0,t), \qquad g_2(t) = u_{xx}(0, t).$$ 
Suppose that the initial and boundary values are compatible in the sense that 
\begin{itemize}
\item they are compatible with equation (\ref{kdv}) at $x=t=0$.
\item the spectral functions satisfy the global relation (\ref{globalrelation}).\footnote{If $T < \infty$ we require that the function $c^+(\zeta)$ on the right-hand side of (\ref{globalrelation}) be continuous and bounded for $\zeta \in \bar{D}_1 \cup \bar{D}_2$, analytic in $D_1 \cup D_2$, and $c^+(k) = O(1/k)$ as $k \to \infty$.}
\end{itemize}
Assume that the possible zeros $\{k_j\}_{1}^{N}$ of $a(k)$ and $\{\lambda_j\}_{1}^{\Lambda}$ of $d(k)$ are as in Assumption \ref{zerosassumption}. Define $M(x,t,k)$ as the solution of the following $2\times 2$ matrix RH problem:
\begin{itemize}
\item $M$ is sectionally meromorphic away from the boundaries of the $D_j$'s, $j = 1, \dots, 4$.

\item $M$ satisfies the jump condition
$$M_-(x,t,k) = M_+(x,t,k)J(x,t,k), \qquad k \in \bar{D}_i \cap \bar{D}_j, \quad i,j = 1, \dots, 4,$$
where $M$ is $M_-$ for $k \in D_2 \cup D_4$, $M$ is $M_+$ for $k \in D_1 \cup D_3$, and $J$ is defined in terms of $a,b,A$, and $B$ by equations (\ref{ddef})-(\ref{Gammadef}).

\item The first column of $M$ has simple poles at $k = k_j$, $j = 1, \dots, n_1$, and $k = \lambda_j$, $j = 1, \dots, \Lambda$. The second column of $M$ has simple poles at $k = \bar{k}_j$ and $k = \bar{\lambda}_j$, $j = 1, \dots, \Lambda$. The associated residues satisfy the following relations:
\begin{align}\label{residue1}
\underset{k_j}{\text{\upshape Res}} [M(x,t,k)]_1 =& \frac{1}{\dot{a}(k_j)b(k_j)} e^{2i\theta(k_j)} [M(x,t,k_j)]_2, \qquad j = 1, \dots, 2n_1,
		\\\label{residue2}
\underset{\bar{k}_j}{\text{\upshape Res}} [M(x,t,k)]_2 =& \frac{1}{\overline{\dot{a}(k_j)b(k_j)}} e^{-2i\theta(\bar{k}_j)} [M(x,t,\bar{k}_j)]_1, \qquad j = 1, \dots, 2n_1,
	\\ \label{residue3}
\underset{\lambda_j}{\text{\upshape Res}} [M(x,t,k)]_1 = &
\underset{\lambda_j}{\text{\upshape Res}} \, \Gamma(k) \, e^{2i\theta(\lambda_j)} [M(x,t,\lambda_j)]_2, \qquad j = 1, \dots, 2\Lambda,
	\\\label{residue4}
\underset{\bar{\lambda}_j}{\text{\upshape Res}} [M(x,t,k)]_2 =& 
\underset{\bar{\lambda}_j}{\text{\upshape Res}} \, \overline{\Gamma(\bar{k})} \,
e^{-2i\theta(\bar{\lambda}_j)} [M(x,t,\bar{\lambda}_j)]_1, \qquad j = 1, \dots, 2\Lambda,
\end{align}
where $\theta(k_j) = -k_j x + (k_j +4k_j^3)t$, and $\text{Res}_{\bar{\lambda}_j} \, \overline{\Gamma(\bar{k})}$ denotes the residue of the function $k \mapsto \overline{\Gamma(\bar{k})}$ at $k = \bar{\lambda}_j$.

\item $M(x,t,k) = I + O\left(\frac{1}{k}\right), \qquad k \to \infty.$

\item For some real function $\alpha(x, t)$, 
$$M(x, t, k) \sim \frac{i \alpha(x, t)}{k}\begin{pmatrix} 0 & 1 \\ 0 & -1 \end{pmatrix}, \qquad k \to 0, \, k \in D_1.$$

\end{itemize}
Then $M(x,t,k)$ exists and is unique.

Define $u(x,t)$ in terms of $M(x,t,k)$ by
\begin{equation}\label{recoveru}
  u(x, t) = -2 i \lim_{k \to \infty} k \partial_x\left( M_{22}(x,t,k)\right).
\end{equation}

Then $u(x,t)$ solves equation (\ref{kdv}). Furthermore,
\begin{equation*}
  u(x,0) = u_0(x), \quad u(0,t) = g_0(t), \quad u_x(0,t) = g_1(t), \quad \text{and} \quad u_{xx}(0,t) = g_2(t).
\end{equation*}
\end{theorem}

Let us also point out that Remark \ref{regularizedremark} applies also to the RH problem for the half-line problem: In order to find the solution $M$, we may first find the solution $\check{M}$ of the regularized RH problem with no singularity at $k = 0$, and then recover $M$ from equation (\ref{MfromregularM}).

\section{Linearizable boundary conditions}\label{linearizablesec} \nequation
It was shown in Theorem \ref{RHtheorem} that the solution $u(x,t)$ of equation (\ref{kdv}) on the half-line can be expressed through the solution of a $2 \times 2$ matrix RH problem, which is uniquely formulated in terms of the spectral functions $a(k)$, $b(k)$, $A(k)$, and $B(k)$. The functions $a(k)$ and $b(k)$ are defined in terms of the initial data $u_0(x)$ through the solution of the linear Volterra integral equation (\ref{mu3Volterra}). However, the spectral functions $A(k)$ and $B(k)$ are, in general, not as readily obtained: The construction of $A(k)$ and $B(k)$ via the linear Volterra integral equation (\ref{mu1Volterra}) requires knowledge of $g_0(t)$, $g_1(t)$, and $g_2(t)$, whereas the given boundary conditions impose only two conditions among these three functions; the additional condition needed to determine $g_0(t)$, $g_1(t)$, and $g_2(t)$ is the requirement that they satisfy the global relation (\ref{globalrelation}). In general, this problem involves solving a nonlinear Volterra integral equation. 

However, for a particular class of boundary value problems it is possible, using only 
{\it the algebraic manipulation of the global relation,} to compute functions $\tilde{A}(k)$ and $\tilde{B}(k)$ which effectively replace $A(k)$ and $B(k)$. More precisely, the solution $\tilde{M}$ to the RH problem associated with $\tilde{A}, \tilde{B}$ instead of $A, B$, can be directly related to the solution of the original RH problem (see Theorem \ref{linearizableTh} below). In particular, $u(x,t)$ can be recovered from the large $k$ asymptotics of $\tilde{M}$. When $T = \infty$ the functions $\tilde{A}$ and $\tilde{B}$ coincide with $A$ and $B$, so that the two RH problems are identical. 
The class of boundary value problems which yield to this approach are referred to as linearizable. Thus, for linearizable IBV problems the half-line formalism is as effective as the formalism on the line. An analysis of linearizable boundary conditions for equation (\ref{kdv}) was outlined in \cite{Fbook}. However, our consideration of specific examples will require some additional information in order to keep track of specific branches of solutions to third-order polynomials; we will therefore in this section present a more detailed version of the analysis of \cite{Fbook}.

The derivation involves considering for each $k$ the roots $\nu(k)$ of the polynomial
\begin{equation}\label{nukrelation}  
  \nu(k) + 4\nu(k)^3 = k + 4k^3.
\end{equation}
For each $k$, there are three roots $\nu_j(k) \in \C$, $j = 1,2,3$ (one of these roots equals $k$), but it is impossible to choose a consistent numbering of these roots on the whole complex plane. Indeed, the two roots not equal to $k$ get interchanged as $k$ encircles one of the points $k = \pm i/2$. However, it is possible to fix a numbering of the roots for $k$ lying in the {\it restricted} set $\bar{D}_1 \cup \bar{D}_3$. For $k \in \bar{D}_1 \cup \bar{D}_3$ one root lies in $\bar{D}_1$ while the two other roots lie in $\bar{D}_3$. We may thus define continuous functions $\nu_1:\bar{D}_1 \cup \bar{D}_3 \to \bar{D}_1$ and $\nu_j:\bar{D}_1 \cup \bar{D}_3 \to \bar{D}_3$, $j = 2,3$, such that
$$\nu_j(k) + 4\nu_j(k)^3 = k + 4k^3, \qquad k \in \bar{D}_1 \cup \bar{D}_3, \quad j = 1,2,3.$$
For each $k \in \bar{D}_1 \cup \bar{D}_3$ exactly one of the complex numbers $\nu_j(k)$, $j = 1,2,3$, equals $k$.
We extend the domain of definition of $\nu_j$, $j = 1,2,3$, to include $D_2 \cup D_4$ by defining
$$\nu_j(k) = \overline{\nu_j(\bar{k})}, \qquad k \in D_2 \cup D_4.$$
With this notation, the basic result is the following.

\begin{theorem}\label{linearizableTh}
Let $u(x,t)$ satisfy equation (\ref{kdv}) on $\{x > 0, 0<t<T \}$ with $T \leq \infty$ together with the initial condition 
$$u(x,0) = u_0(x), \qquad 0 < x < \infty,$$
and one of the following boundary conditions:
\begin{itemize}
\item[(a)] $u(0,t) = u_{xx}(0,t) = 0,$

\item[(b)] 
$u(0,t) = \chi, \quad u_{xx}(0,t) = \chi + 3\chi^2, \quad \chi \in \R, \quad \chi \neq 0.$
\end{itemize}
In the case (a) we define
\begin{align} \label{aDeltadef}
\Delta(k) = & a(k) \frac{\overline{a(\nu_1(\bar{k}))}}{\overline{b(\nu_1(\bar{k}))}} - b(k) , \qquad k \in \bar{D}_2,
	\\ \label{aGammatilde}
\tilde{\Gamma}(k) = & \frac{\overline{b(\nu_1(\bar{k})})}{a(k) \Delta(k)}, \qquad k \in \bar{D}_2.
\end{align}
In the case (b) we define
\begin{align} \label{bf1def}
&f_1(k) = \frac{\nu_1(k) + k}{\nu_1(k) - k}\left(1 - \frac{4k\nu_1(k)}{\chi}\right),
	\\ \label{bDeltadef}
&\Delta(k) = a(k) \frac{f_1(k)\overline{a(\nu_1(\bar{k}))} - \overline{b(\nu_1(\bar{k}))}}{f_1(k)\overline{b(\nu_1(\bar{k}))} - \overline{a(\nu_1(\bar{k}))}} - b(k), \quad k \in \bar{D}_2,
 	\\ \label{bGammatilde}
&\tilde{\Gamma}(k) = \frac{1}{a(k) \Delta(k)}, \qquad k \in \bar{D}_2.
\end{align}
Assume that the initial and boundary conditions are compatible at $(x,t) = (0,0)$. If $T= \infty$, we assume that the boundary conditions are of type (a). Furthermore, assume that the following are valid:
\begin{itemize}
\item[(i)] $a(k)$ has $N$ simple zeros $\{k_j\}_{j = 1}^{N} \subset i \R^-$ such that $k_j \in D_1$, $j = 1, \dots, n_1$, and $k_j \in D_2$, $j = n_1+1, \dots, N$. 

\item[(ii)] $\Delta(k)$ has $\Lambda$ simple zeros $\{\lambda_j\}_1^{\Lambda}  \subset i \R^-$ such that $\lambda_j \in D_2$, $j = 1, \dots, \Lambda$.

\item[(iii)] None of the zeros of $a(k)$ coincides with a zero of $\Delta(k)$. 
\end{itemize}

Then the solution $u(x,t)$ is given by equation (\ref{recoveru}) with $M$ replaced by $\tilde{M}$, where $\tilde{M}$ is the solution of the Riemann-Hilbert problem in Theorem \ref{RHtheorem} with jump matrices and residue conditions defined by replacing $\Gamma(k)$ in (\ref{J123def}) and (\ref{residue3})-(\ref{residue4}) with $\tilde{\Gamma}(k)$. 

Let $\tilde{J}_i$, $i = 1, \dots, 4$, be the jump matrices defined according to (\ref{J123def}) with $\Gamma$ replaced by $\tilde{\Gamma}$. Then $\tilde{M}$ is related to the solution $M$ of the original RH problem by
\begin{equation}\label{MtildeMrelations}
M_1 = \tilde{M}_1, \quad 
M_2 = \tilde{M}_2 \tilde{J}_1^{-1} J_1, \quad 
M_3 = \tilde{M}_3 \tilde{J}_3 J_3^{-1}, \quad 
M_4 = \tilde{M}_4.
\end{equation}
Moreover, if $T = \infty$,
\begin{equation}\label{tildeGammaGamma}
  \tilde{\Gamma}(k) = \Gamma(k),
\end{equation}
so that the two RH problems coincide.
\end{theorem}

\proofbegin
Suppose that it is possible to compute explicitly in terms of the given boundary conditions, a nonsingular matrix $N_j(k)$ such that, for $ j = 1,2,3$ and $k \in \C$,
\begin{equation}\label{Nintertwine}
\left[i(k + 4k^3) \sigma_3 - V_2(0,t,\nu_j(k))\right]N_j(k) = N_j(k) \left[i(k + 4k^3) \sigma_3 - V_2(0,t,k)\right].
\end{equation}
If $N_j(k)$ exists and $T < \infty$, it follows from (\ref{lax}), (\ref{Seq}), and (\ref{Nintertwine}) that
\begin{equation}\label{SnuNSNinv}
S(\nu_j(k)) = N_j(k) S(k) e^{i (k + 4k^3) T \hat{\sigma}_3} N_j^{-1}(k), \qquad k \in \C.
\end{equation}
It turns out that matrices $N_j(k)$ satisfying (\ref{Nintertwine}) exist for the boundary conditions in (a) and (b). 

We can now explain the idea of the proof: The spectral functions $A(k)$ and $B(k)$ enter the formulation of the RH problem in Theorem \ref{RHtheorem} only through the function $\Gamma(k)$ defined in (\ref{Gammadef}). The goal is to use the algebraic relations between $S(\nu_j(k))$ and $S(k)$ obtained from (\ref{SnuNSNinv}) together with the global relation (\ref{globalrelation}) to express $\Gamma(k)$ in terms of $a$, $b$, and the matrices $N_j$. This turns out to be possible when $T = \infty$. When $T < \infty$ we will define modified spectral functions $\tilde{A}(k)$ and $\tilde{B}(k)$ and define $\tilde{\Gamma}$ by the same equation (\ref{Gammadef}) as we used to define $\Gamma$, but with $A$ and $B$ replaced by $\tilde{A}$ and $\tilde{B}$, i.e.
\begin{equation}\label{Gammatildedef}
  \tilde{\Gamma}(k) = \frac{\overline{B(\bar{k})}}{a(k)(a(k)\overline{A(\bar{k})} -  b(k)\overline{B(\bar{k})})}, \quad k \in \bar{D}_2.
\end{equation}
It can then be shown that the solutions of the RH problems formulated in terms of $\Gamma$ or $\tilde{\Gamma}$ are related as in (\ref{MtildeMrelations}). 

The first step is to show that $\tilde{\Gamma}(k)$ as defined in (\ref{Gammatildedef}) can be expressed as in (\ref{aGammatilde}) and (\ref{bGammatilde}) when $u$ satisfies boundary conditions of type (a) and type (b), respectively. We consider the two cases below.
 

\subsection{Boundary conditions of type (a)}
Assume first that $T < \infty$. In this case (\ref{Nintertwine}) holds with $N_j = I$, $j = 1,2,3$, which using (\ref{SnuNSNinv}) leads to
\begin{equation}\label{ABnuj}
  A(\nu_j(k)) = A(k), \qquad B(\nu_j(k)) = B(k), \qquad j = 1,2,3.
\end{equation}
In view of this symmetry, the global relation (\ref{globalrelation}) yields
\begin{equation}\label{globalnuj}
B(k) a(\nu_1(k)) - A(k) b(\nu_1(k)) = e^{2i (k + 4k^3) T} c^+(\nu_1(k)), \qquad k \in \bar{D}_1 \cup \bar{D}_3.
\end{equation}
Moreover, (\ref{ABnuj}) together with the definition of $d$,
\begin{equation}\label{ddefrepeat}
  \overline{d(\bar{k})} = \overline{a(\bar{k})} A(k) - \overline{b(\bar{k})} B(k), \qquad k \in \bar{D}_3,
\end{equation}  
imply
\begin{equation}\label{dnuj}
\overline{d(\overline{\nu_j(k)})} = \overline{a(\overline{\nu_j(k)})} A(k) - \overline{b(\overline{\nu_j(k)})} B(k), \qquad k \in \bar{D}_1 \cup \bar{D}_3, \quad j = 2,3.
\end{equation}
Equations (\ref{globalnuj}) and (\ref{dnuj}) can be solved for $A$ and $B$ to give, for $j = 2,3$ and $k \in \bar{D}_1 \cup \bar{D}_3$, the following expressions:
\begin{align} \label{Acasea}
A(k) = \frac{a(\nu_1(k))\overline{d(\overline{\nu_j(k)})} + e^{2i (k + 4k^3) T} c^+(\nu_1(k))\overline{b(\overline{\nu_j(k)})}}{\Delta_j(k)},
	\\  \label{Bcasea}
B(k) = \frac{b(\nu_1(k))\overline{d(\overline{\nu_j(k)})} + e^{2i (k + 4k^3) T} c^+(\nu_1(k))\overline{a(\overline{\nu_j(k)})}}{\Delta_j(k)},
\end{align}
where
$$\Delta_j(k) = a(\nu_1(k))  \overline{a(\overline{\nu_j(k)})} - b(\nu_1(k))  \overline{b(\overline{\nu_j(k)})}, \qquad k \in \bar{D}_1 \cup \bar{D}_3.$$
The last terms on the right-hand sides of equations (\ref{Acasea}) and (\ref{Bcasea}) involve the unknown function $c^+$. We are therefore led to consider the RH problem with $A$ and $B$ replaced by 
\begin{align}
\tilde{A}(k) = \frac{a(\nu_1(k))\overline{d(\overline{\nu_j(k)})}}{\Delta_j(k)},
	\\
\tilde{B}(k) = \frac{b(\nu_1(k))\overline{d(\overline{\nu_j(k)})}}{\Delta_j(k)}.
\end{align}
Let $\tilde{\Gamma}$ be defined by (\ref{Gammatildedef}), thus
$$\tilde{\Gamma}(k) = \frac{\overline{\tilde{B}(\bar{k})}/\overline{\tilde{A}(\bar{k})}}{a(k) \left[a(k) - b(k) \overline{\tilde{B}(\bar{k})}/\overline{\tilde{A}(\bar{k})}\right]} 
= \frac{1}{a(k) \Delta(k)}, \qquad k \in \bar{D}_2,$$
where $\Delta(k)$ is given by (\ref{aDeltadef}).

For $T = \infty$ the derivation is similar, but simpler. In particular, $\tilde{A}$ and $\tilde{B}$ coincide with $A$ and $B$ in this case.
This establishes formula (\ref{aGammatilde}) for $\tilde{\Gamma}$ in case (a).

\subsection{Boundary conditions of type (b)}
Let 
$$f_j(k) =\frac{\nu_j(k) + k}{\nu_j(k) - k}\left(1 - \frac{4k\nu_j(k)}{\chi}\right),\qquad j = 1,2,3.$$
Then (\ref{Nintertwine}) is satisfied for 
$$N_j(k) = \begin{pmatrix} f_j(k) &	1 \\
	1 & f_j(k) \end{pmatrix}.$$
Equation (\ref{SnuNSNinv}) gives, for $k \in \C$ (we again assume $T < \infty$),
\begin{align}\label{Anu}
(f_j(k)^2 - 1)A(\nu_j(k)) = f_j(k) \left[B(k) + f_j(k) A(k)\right]  - e^{2i (k + 4k^3) T} \left[\overline{A(\bar{k})} + f_j(k) \overline{B(\bar{k})}\right],
	\\ \label{Bnu}
(f_j(k)^2 - 1)B(\nu_j(k)) = f_j(k) \left[A(k) + f_j(k) B(k)\right]  - e^{2i (k + 4k^3) T} \left[\overline{B(\bar{k})} + f_j(k) \overline{A(\bar{k})}\right].
\end{align}
Note that $f_j(k) = \overline{f_j(\bar{k})}$, $j = 1,2,3$.
By taking the Schwarz conjugate of (\ref{Anu})-(\ref{Bnu}) we find, for $k \in \C$,
\begin{align} \label{schwartzAnu}
(f_j(k)^2 - 1)\overline{A(\nu_j(\bar{k}))} = f_j(k) \left[\overline{B(\bar{k})} + f_j(k)\overline{A(\bar{k})}\right]  - e^{-2i (k + 4k^3) T} \left[A(k) + f_j(k)B(k)\right],
	\\\label{schwartzBnu}
(f_j(k)^2 - 1)\overline{B(\nu_j(\bar{k}))} = f_j(k) \left[\overline{A(\bar{k})} + f_j(k) \overline{B(\bar{k})}\right]  - e^{-2i(k + 4k^3) T} \left[B(k) + f_j(k)A(k)\right].
\end{align}
Eliminating $\overline{B(\bar{k})} + f_j(k) \overline{A(\bar{k})}$ from equations (\ref{Bnu}) and (\ref{schwartzAnu}), we find
\begin{equation}\label{Bnujk}
B(\nu_j(k)) = - \frac{e^{2i(k + 4k^3) T}\overline{A(\nu_j(\bar{k}))}}{f_j(k)} + \frac{A(k)}{f_j(k)} + B(k), \qquad k \in \C, \quad j = 1,2,3.
\end{equation}
Similarly, eliminating $\overline{A(\bar{k})} + f_j(k) \overline{B(\bar{k})}$ from equations (\ref{Anu}) and (\ref{schwartzBnu}), we find
\begin{equation}\label{Anujk}
A(\nu_j(k)) = - \frac{e^{2i(k + 4k^3) T}\overline{B(\nu_j(\bar{k}))}}{f_j(k)} + \frac{B(k)}{f_j(k)} + A(k), \qquad k \in \C, \quad j = 1,2,3.
\end{equation}

In order to obtain $A(k)$ and $B(k)$ for $k \in \bar{D}_3$ we replace $A(\nu_1(k))$ and $B(\nu_1(k))$ by the right-hand sides of (\ref{Bnujk}) and (\ref{Anujk}) in the global relation
\begin{equation}\label{globalnuj3}
B(\nu_1(k)) a(\nu_1(k)) - A(\nu_1(k)) b(\nu_1(k)) = e^{2i (k + 4k^3) t} c^+(\nu_1(k)), \qquad k \in \bar{D}_3.
\end{equation}
This gives
$$A(k) \left[ \frac{a(\nu_1(k))}{f_1(k)} - b(\nu_1(k))\right]
- B(k) \left[ \frac{b(\nu_1(k))}{f_1(k)} - a(\nu_1(k))\right]
= e^{2i (k + 4k^3) T}c^+(\nu_1(k)) $$
$$+ \frac{e^{2i (k + 4k^3) T}}{f_1(k)} \left[\overline{A(\overline{\nu_1(k)})} a(\nu_1(k)) - \overline{B(\overline{\nu_1(k)})} b(\nu_1(k))\right], \qquad k \in \bar{D}_3.
$$
Using the definition (\ref{ddef}) of $d$, which is valid for $k \in \bar{D}_3$, we infer that
$$A(k) \left[ \frac{a(\nu_1(k))}{f_1(k)} - b(\nu_1(k))\right]
- B(k) \left[ \frac{b(\nu_1(k))}{f_1(k)} - a(\nu_1(k))\right]$$
$$= e^{2i (k + 4k^3) T}\left[c^+(\nu_1(k)) + \frac{d(\nu_1(k))}{f_1(k)}\right], \qquad k \in \bar{D}_3.
$$
Solving this equation together with (\ref{ddefrepeat}) for $A$ and $B$, we find
\begin{align} \label{AfinalD3}
  A(k) = &\frac{1}{\overline{\Delta_1(\bar{k})}} \biggl[ \left( a(\nu_1(k)) - \frac{b(\nu_1(k))}{f_1(k)} \right) \overline{d(\bar{k})} 
  		\\ \nonumber
 & + e^{2i (k + 4k^3) T} \overline{b(\bar{k})}\left(c^+(\nu_1(k)) + \frac{d(\nu_1(k))}{f_1(k)}\right) \biggr], \qquad k \in \bar{D}_3,   	\\ \label{BfinalD3}
  B(k) =& \frac{1}{\overline{\Delta_1(\bar{k})}}  \biggl[ \left( b(\nu_1(k)) - \frac{a(\nu_1(k))}{f_1(k)} \right) \overline{d(\bar{k})} 
  	\\ \nonumber
 & + e^{2i (k + 4k^3) T} \overline{a(\bar{k})}\left(c^+(\nu_1(k)) + \frac{d(\nu_1(k))}{f_1(k)} \right) \biggr], \qquad k \in \bar{D}_3,
  \end{align}
where, for $k \in \bar{D}_2$,
$$\Delta_1(k) = a(k)\overline{a(\overline{\nu_1(k)})} - b(k)\overline{b(\overline{\nu_1(k)})} + \frac{1}{f_1(k)}\left[-a(k)\overline{b(\overline{\nu_1(k)})} + b(k) \overline{a(\overline{\nu_1(k)})}\right].$$
Some of the terms of the right-hand sides of equations (\ref{AfinalD3}) and (\ref{BfinalD3}) involve unknown functions, so that we are led to consider the RH problem for which $A$ and $B$ are replaced by the following expressions:
\begin{align}
  &\tilde{A}(k) = \frac{1}{\overline{\Delta_1(\bar{k})}}  \left[ a(\nu_1(k)) - \frac{b(\nu_1(k))}{f_1(k)} \right] \overline{d(\bar{k})} ,  \qquad k \in \bar{D}_3, 	\\
 & \tilde{B}(k) = \frac{1}{\overline{\Delta_1(\bar{k})}}  \left[ b(\nu_1(k)) - \frac{a(\nu_1(k))}{f_1(k)} \right] \overline{d(\bar{k})} , \qquad k \in \bar{D}_3.
  \end{align}
Then
\begin{align*}
\tilde{\Gamma}(k) &\, = \frac{\overline{\tilde{B}(\bar{k})}/\overline{\tilde{A}(\bar{k})}}{a(k) \left[a(k) - b(k) \overline{\tilde{B}(\bar{k})}/\overline{\tilde{A}(\bar{k})}\right]}
	\\
& = \frac{1}{a(k) \Delta(k)}, \qquad k \in \bar{D}_2,
\end{align*}
where $\Delta(k)$ is given by (\ref{bDeltadef}).
This establishes formula (\ref{bGammatilde}) for $\tilde{\Gamma}$ in case (b).

\bigskip
It remains to show that the solutions $\tilde{M}$ and $M$ of the RH problems defined in terms of $\tilde{\Gamma}$ and $\Gamma$ are related by (\ref{MtildeMrelations}), and that $\Gamma(k) = \tilde{\Gamma}(k)$ when $T = \infty$. It will then follow from (\ref{MtildeMrelations}) that $u(x,t)$ is given by equation (\ref{recoveru}) with $M$ replaced by $\tilde{M}$. The arguments necessary to prove (\ref{MtildeMrelations}) are presented in \cite{F2002}. The equality $\Gamma(k) = \tilde{\Gamma}(k)$ when $T = \infty$ is a consequence of the fact that $\tilde{A} = A$ and $\tilde{B} = A$ in this case. The restriction to case (a) when $T = \infty$ is necessary in order for the eigenfunction $\mu_1$ (and thus also for $\Gamma$) to be well-defined, see subsection \ref{boundedanalyticsubsec}.
This completes the proof of Theorem \ref{linearizableTh}.
$\hfill\Box$

\begin{remark} \label{conservedremark}\upshape {\bf (Characterization of linearizable boundary conditions)}
The linearizable boundary conditions can be characterized as those boundary  conditions which are time-independent and for which the total mass $\int_0^\infty u dx$ is conserved. This follows directly from the fact that
\begin{equation}\label{ddtint0infty}
  \frac{d}{dt} \int_0^\infty u(x,t) dx =  \bigl[u_{xx} - u - 3u^2\bigr]_{x = 0}.
\end{equation}
Equation (\ref{ddtint0infty}) is a consequence of the conservation law
\begin{equation}\label{conservedmasslaw}
  u_t + \left(u - u_{xx} + 3u^2\right)_x = 0.
\end{equation}
\end{remark}

\section{Soliton profiles}\label{solitonsec} \nequation
We now begin the study of particular examples of IBV problems. The first examples are derived from soliton solutions of equation (\ref{kdv}). The expressions for these solutions and the corresponding eigenfunctions of the Lax pair can be found from the RH problem in Theorem \ref{lineRHtheorem} as indicated in subsection \ref{solitonsubsec}. We will consider the one-soliton and the two-solitons, and will also comment on the case of rational solitons.

\subsection{One-solitons}\label{onesolitonsubsec} 
\begin{figure}
\begin{center}
    \includegraphics[width=.6\textwidth]{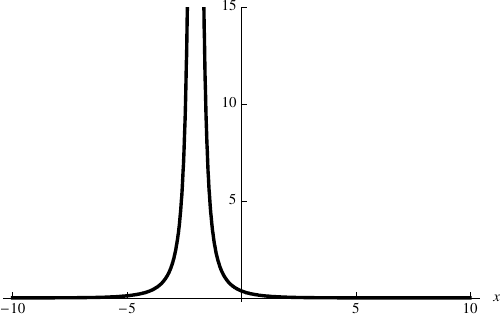}
     \begin{figuretext}\label{cschsolp-05x0-1.pdf}
        The $\text{csch}^2$-soliton $u_p^+(x,t)$ for $p = -1/2$ and $x_0 = -1$ at time $t = 0$.
     \end{figuretext}
     \end{center}
\end{figure}

In order to find the one-soliton of equation (\ref{kdv}) and the corresponding eigenfunctions, we consider the algebraic system (\ref{algebraicsystem}) with $N=2$ and $k_2 = -\bar{k}_1$ (in view of the symmetries (\ref{musymmetries}), if $a^L(k) = 0$ then also $a^L(-\bar{k}) = 0$). 
Since $\theta_j = -k_j x + (k_j +4k_j^3)t$, we have $\theta_1 = - \bar{\theta}_2$. The identity $\overline{a^L(-\bar{k})} = a^L(k)$ implies that $\dot{a}^L(k_2) = -\overline{\dot{a}^L(k_1)}$. Moreover, (\ref{linemubj}) together with the symmetries (\ref{musymmetries}) imply that $\bar{b}_2 = b_1$. Hence,
$C_2 = -\bar{C}_1$ and, using (\ref{Msymmetry}), the system (\ref{algebraicsystem}) can be written as
\begin{equation}\label{solitonalgebraicsystem}
\begin{pmatrix} \overline{\check{M}_{22}(k_1)} \\ \overline{\check{M}_{12}(k_1)} \end{pmatrix} = \begin{pmatrix} 1	\\	0 \end{pmatrix} 
+ \frac{C_1 e^{2i\theta_1}}{\bar{k}_1 - k_1}\begin{pmatrix} \check{M}_{12}(k_1) \\ \check{M}_{22}(k_1) \end{pmatrix}
 - \frac{\bar{C}_1 e^{-2i\bar{\theta}_1}}{\bar{k}_1 + \bar{k}_1}
\begin{pmatrix} \overline{\check{M}_{12}(k_1)} \\ \overline{\check{M}_{22}(k_1)} \end{pmatrix}.
\end{equation}
Since we consider only real-valued solutions, the zeros of $a^L$ will lie on the imaginary axis, so we may assume that $k_1 = ip = k_2$ where $p > 0$. We parametrize the normalization constant $C_1$ as $C_1 = -i \sigma p e^{2 x_0}$, where $\sigma = \pm 1$ and $x_0 \in \R$. From (\ref{Msymmetry}) we deduce that $\check{M}_{12}(k_1)$ and $\check{M}_{22}(k_1)$ are real. Substituting the solution of (\ref{solitonalgebraicsystem}) into (\ref{M2kfromM2kj}), we obtain 
\begin{align*}
& \check{M}_{12}(k) = -\frac{2 \sigma e^{2 \left(4 t p^3+(x-t) p+x_0\right)} p}{\left(-1+e^{4
   \left(4 t p^3+(x-t) p+x_0\right)}\right) (p-i k)},
   	\\
& \check{M}_{22}(k) = 1-\frac{2 e^{4 \left(4 t p^3+x p+x_0\right)} p}{\left(-e^{4 p
   t}+e^{4 \left(4 t p^3+x p+x_0\right)}\right) (p-i k)}.
\end{align*}   
Hence, the function $y(x)$ in (\ref{MfromregularM}) is given by
$$y(x) = \frac{4 \sigma e^{2 \left(4 t p^3+(x-t) p+x_0\right)} p}{-1+e^{4 \left(4
   t p^3+(x-t) p+x_0\right)}}.$$
We can now compute $M$ according to (\ref{MfromregularM}) and then find the one-soliton $u_{1}(x,t)$ from (\ref{linerecoveru}). By (\ref{MplusMminus}), the second column of $M$ equals $\mu_3^{(12)}$ for $k \in D_1 \cup D_2$. Letting $\sigma = -1$ we find
\begin{align}\label{onesoliton}  
 & u_{p}^-(x,t) = -2 p^2 \text{sech}^2\left(4 t p^3+(x-t) p + x_0\right),
	\\ \label{onesolitonmu3}  
& [\mu_3(x,t,k)]_2 = \left(
\begin{array}{l}
 -\frac{p^2 \text{sech}^2\left(4 t p^3+(x-t) p+x_0\right)}{2 k (k+i
   p)} \\
\frac{e^{4 p t} k (p-i k)-e^{4 \left(4 t p^3+x p+x_0\right)} k (i
   k+p)-2 i e^{2 \left(4 t p^3+(t+x) p+x_0\right)}
   \left(k^2+p^2\right)}{\left(e^{2 p t}+e^{2 \left(4 t p^3+x
   p+x_0\right)}\right)^2 k (p-i k)}
\end{array}
\right).
\end{align}
The results for $\sigma = 1$ are obtained from these expressions by shifting $x_0 \to x_0 + i \pi/2$. This gives the following additional class of solitons, see Figure \ref{cschsolp-05x0-1.pdf}:
\begin{equation}\label{sinhonesoliton}  
  u_{p}^+(x,t) = 2 p^2 \text{csch}^2\left[4 p^3 t + p (-t + x) + x_0\right].
\end{equation}

\begin{figure}
\begin{center}
    \includegraphics[width=.6\textwidth]{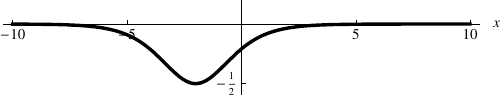}
     \begin{figuretext}\label{sechsolp-05x0-1.pdf}
        The $\text{sech}^2$-soliton $u_p^-(x,t)$ for $p = -1/2$ and $x_0 = -1$ at time $t = 0$.
     \end{figuretext}
     \end{center}
\end{figure}

The solution $u_{p}^-(x,t)$ is the well-known $\text{sech}^2$-soliton, see Figure \ref{sechsolp-05x0-1.pdf}. For KdV this soliton has positive height, but for equation (\ref{kdv}) it has the opposite sign ($u_p^-(x,t) < 0$ for all $x, t$). The solution $u_{p}^+(x,t)$ is less standard; it is normally not considered, because for each time $t$ it has a singularity at 
$$x = \frac{-4 t p^3+t p-x_0}{p}.$$
However, for the half-line problem $u_{p}^+(x,t)$ is a perfectly smooth solution for all $t > 0$ whenever $p \leq - 1/2$ and $x_0 < 0$, in which case the singularity never enters the positive half of the $x$-axis.

\subsubsection{Asymptotic solitons}
Assume $T = \infty$ and let $u(x,t)$ be the solution of an IBV problem on the half-line. Assume that $u(0,t)$ decays to zero as $t \to \infty$. An application of the nonlinear steepest descent method of \cite{D-Z} to the RH problem of Theorem \ref{RHtheorem} shows that the solution $u(x,t)$ will asymptote for large $t$ into a sum of one-solitons traveling at constant speeds; these solitons are generated by the poles of $a(k)\Gamma(k)$. Away from these solitons the solution is of dispersive character (cf. \cite{FI94}), i.e.
\begin{equation}\label{usumsolitons} 
  u(x,t) = \sum_{n} u_{p_n}(x,t) + o(1), \qquad t\to \infty, \quad \frac{x}{t} = O(1),
\end{equation}
where we have denoted the poles by $\{ip_n\}_1^\Lambda$, $-1/2 \leq p_1 < p_2 < \cdots < p_\Lambda < 0$. In view of the assumption \ref{zerosassumption}, the set of poles of $a(k)\Gamma(k)$ coincides with the set of zeros of $d(k)$, unless a zero of $d(k)$ happens to coincide with a zero of $\overline{B(\bar{k})}$; therefore we usually say that the solitons are generated by the zeros of $d(k)$ even though strictly speaking some of these zeros may cancel with zeros of $\overline{B(\bar{k})}$. 

If $u(x,t)$ satisfies a linearizable boundary condition at $x = 0$, then a similar asymptotic analysis of the RH problem for $\tilde{M}$ in Theorem \ref{linearizableTh} shows that formula (\ref{usumsolitons}) holds also in this case with $\{ip_n\}_1^\Lambda$, $-1/2 \leq p_1 < p_2 < \cdots < p_ \Lambda < 0$, denoting the poles of $a(k)\tilde{\Gamma}(k) = 1/\Delta(k)$. We draw the following important conclusion:
\begin{quote}{\it
For a linearizable IBV problem the asymptotic solitons are generated by the zeros of $\Delta(k)$.}
\end{quote}

Which solitons can be generated asymptotically for an IBV problem on the half-line? We first note that $-1/2 \leq p_1 < p_2 < \cdots < p_ \Lambda < 0$. Hence, whereas equation (\ref{kdv}) admits solitons $u_p^\pm(x,t)$ for any value of $p < 0$, only those satisfying $-\frac{1}{2} \leq p < 0$ can be generated asymptotically for the half-line problem. Moreover, among the solitons with $-\frac{1}{2} \leq p < 0$, the solitons of $\text{csch}^2$-type are singular unless $p = -1/2$. We conclude that the solitons that can be generated asymptotically are: (a) the $\text{sech}^2$-solitons $u_p^-(x,t)$ for $-\frac{1}{2} \leq p < 0$ and (b) the $\text{csch}^2$-solitons $u_p^-(x,t)$ for $p = -\frac{1}{2}$.
The interpretation of the requirement $p \geq -1/2$ is clear: the soliton $u_p^\pm(x,t)$ travels at speed $1 - 4p^2$, so that $p < -\frac{1}{2}$ if and only if the soliton travels to the left; on the half-line these left-moving solitons cannot appear asymptotically.

\subsubsection{Spectral functions}
Using the expression (\ref{onesolitonmu3}) for $\mu_3$, we can derive explicit formulas for the associated spectral functions.
We will consider the case of $u_p^-(x,t)$; the analogous quantities for $u_p^+(x,t)$ can be obtained by shifting $x_0 \to x_0 + i \pi/2$. 
First note that 
$$\mu_3(x,t,k) = \frac{i p \text{sech}^2\left[4 t p^3+(x-t) p+x_0\right]}{2 k}\begin{pmatrix} 1 & 1 \\ -1 & -1 \end{pmatrix} + O(1),\qquad k \to 0,$$
in accordance with (\ref{muatorigin}). Evaluating formula (\ref{onesolitonmu3}) for $\mu_3$ at $x = t = 0$, we find 
\begin{equation}\label{abonesol}
a(k) = \frac{2 k^2-2 i p \tanh (x_0) k+p^2 \text{sech}^2(x_0)}{2
   k^2+2 i p k}, \qquad b(k) = -\frac{p^2 \text{sech}^2(x_0)}{2 k^2+2 i p k}.
 \end{equation}
After computing $s(k)$ and $\mu_3(x,t,k)$ for all $x$ and $t$, we can compute $\mu_2$ by employing equation (\ref{seq}). Then $S(k)$ is found from (\ref{Ssdef}). For a given finite value of $T$, we find the following expressions:
\begin{align} \nonumber
A(k; T)& = \frac{\text{sech}^2\left(-4 T p^3+T p-x_0\right)
   \text{sech}^2(x_0)}{8 k^2 \left(k^2+p^2\right)} 
    \biggl[ \left(k^2-p^2\right) \cosh \left(2 p T-8 p^3 T\right) k^2
   	\\ \nonumber
& +\left(k^2+p^2\right) \cosh \left(-8 T p^3+2 T p-4 x_0\right) k^2
    +2 \bigl(k^4-i p \sinh \left(2 p T-8 p^3 T\right)  k^3
   	\\  \label{ABTonesol}
&-i p \sinh \left(-8 T p^3+2 T p-2 x_0\right) k^3
   	\\ \nonumber
& -i p \sinh (2 x_0) k^3+2 p^2 k^2
+\left(k^2+p^2\right) \cosh (2 x_0)  k^2
   	\\ \nonumber
& +\left(k^2+p^2\right) \cosh \left(2 \left(4 T p^3-T   p+x_0\right)\right) k^2
-i p^3 \sinh \left(-8 T p^3+2 T p-   x_0\right) k
   	\\ \nonumber
& -i p^3 \sinh (2 x_0) k
-e^{2 i \left(4 k^3+k\right) T} p^4+p^4\bigr)\biggr],
   	\\ \nonumber
   B(k; T)& = \frac{p^2 \text{sech}^2\left(-4 T p^3+T p-x_0\right) \text{sech}^2(x_0)} {4 k^2 \left(k^2+p^2\right)} \biggl[e^{2 i \left(4 k^3+k\right) T} k^2
 +e^{2 i \left(4 k^3+k\right) T} \cosh (2 x_0) k^2
   	\\\nonumber
&  -\cosh \left(2   \left(4 T p^3-T p+x_0\right)\right) k^2
 -k^2
 +i p \sinh \left(-8 T p^3+2 T p-2 x_0\right) k
    	\\\nonumber
& +i e^{2 i \left(4 k^3+k\right) T} p \sinh (2 x_0) k
 +e^{2 i \left(4 k^3+k\right) T} p^2-p^2\biggr].
\end{align}   
In the limit $T \to \infty$, for $k \in \bar{D}_1 \cup \bar{D}_3$ and $-1/2 < p < 0$, this yields
\begin{equation}\label{ABonesol}
  A(k; \infty) = \frac{2 k^2-2 i p \tanh (x_0) k+p^2 \text{sech}^2(x_0)}{2
   k^2-2 i k p}, \qquad B(k; \infty) = -\frac{p^2 \text{sech}^2(x_0)}{2 k^2-2 i k p}.
\end{equation}  
As $k \to \infty$, we have $a(k) \to 1$, $b(k) \to 0$, $A(k; T) \to 1$, and $B(k; T) \to 0$ in agreement with the results of subsection \ref{spectralhalflinesubsec}. 
We may also verify directly that
$$a(k) = -\frac{i p}{2 k} + O(1),\qquad b(k) = \frac{i p}{2 k} + O(1),\qquad k \to 0,$$
in accordance with (\ref{abkto0}), and that
$$A(k; T) = \frac{i \beta}{k} + O(1),\qquad 
B(k; T) = -\frac{i \beta}{k} + O(1),\qquad k \to 0,$$
where
   $$\beta = -\frac{1}{4} p 
   \frac{2 p T+\sinh\left(2 p T-8 p^3 T\right)}{\cosh^2\left(p T-4 p^3 T\right) }$$
in accordance with (\ref{ABkto0}).
As another consistency check we note that, if $T = \infty$, it is straightforward to verify the global relation (\ref{globalrelation}).

The spectral functions $a^L(k)$ and $b^L(k)$ on the line can be computed from the formula
$$s^L(k) = \lim_{x\to -\infty} e^{i(-kx + (k + 4k^3)t) \hat{\sigma}_3} \mu_3(x,0,k).$$
Hence
$$\begin{pmatrix} b^L(k) \\ a^L(k) \end{pmatrix} 
= \begin{pmatrix} 
 0
\\ 
 \frac{k-i p}{k+i p}
\end{pmatrix}.$$
For the problem on the line the solitons are generated by the zeros of $a^L(k)$. As expected $a^L(k)$ has a zero at $k = ip$ corresponding to the soliton $u_p^-(x,t)$.
For the half-line problem with $T= \infty$, the solitons are generated by the poles of the function $\Gamma(k)$ defined in (\ref{Gammadef}), or, equivalently, by the zeros of $d(k)$. A computation using (\ref{abonesol}) and (\ref{ABTonesol}) yields
\begin{align*}d(k; T) = \frac{\text{sech}^2\left(-4 T p^3+T p-x_0\right)}{2 k (k+i p)}
 \bigl[&\cosh
   \left(2 \left(4 T p^3-T p+x_0\right)\right) k^2+k^2
   	\\
 &  +i p \sinh
   \left(-8 T p^3+2 T p-2 x_0\right) k+p^2\bigr].
 \end{align*}
Letting $T \to \infty$, we obtain
$$d(k; \infty) = \frac{k-i p}{k+i p}, \qquad k \in \bar{D}_2, \quad -1/2 < p < 0.$$
As expected, $d(k)$ has a zero at $k = ip$ corresponding to $u_p^-(x,t)$.

Finally, the definition (\ref{Gammadef}) together with the above expressions for the spectral functions, yields (we display the result only in the limit $T \to \infty$)
\begin{equation}\label{Gammaonesol}
  \Gamma(k; \infty) = -\frac{(k+i p) p^2}{(k-i p) \left(\cosh (2 x_0) k^2+k^2-i p \sinh
   (2 x_0) k+p^2\right)}, \qquad -1/2 < p < 0.
\end{equation}

\subsubsection{Linearizable one-solitons}\label{linonesolsubsubsec}
The one-solitons $u_p^\pm(x, t)$ satisfy linearizable boundary conditions of type (b) with $\chi$ given as follows:
\begin{equation}\label{chionesol}
\chi = \begin{cases} -\frac{1}{2}\text{sech}^2 x_0, \qquad \sigma = -1, \\ 
\frac{1}{2} \text{csch}^2 x_0, \qquad \sigma = 1,
\end{cases}
\end{equation}
provided that $p$ is given by $p = -1/2$. These are stationary (time-independent) solitons.

Let us compute $\tilde{\Gamma}(k)$ as defined in (\ref{bGammatilde}) when $\sigma = -1$. We first define the function $f_1(k)$ by (\ref{bf1def}) with $\chi = -1/2$ obtained from (\ref{chionesol}). Using the expressions in (\ref{abonesol}) for $a(k)$ and $b(k)$ and recalling that $\nu_1(k)$ satisfies (\ref{nukrelation}), we compute
\begin{equation}\label{f1baquotientonesol}
  \frac{f_1(k) a(\nu_1(k)) - b(\nu_1(k))}{f_1(k) b(\nu_1(k)) - a(\nu_1(k))} = -4 \cosh (2 x_0) k^2-4 k^2-2 i \sinh (2 x_0) k-1.
\end{equation}
Using this in (\ref{bDeltadef}) and (\ref{bGammatilde}) we find
$$\Delta(k) = -4 k (2 k+i) \cosh ^2(x_0),$$
and
\begin{equation}\label{tildeGammaonesol}  
  \tilde{\Gamma}(k) = -\frac{(2 k-i) \text{sech}^2(x_0)}{(2 k+i) \left(8 k^2+4 i \tanh
   (x_0) k+\text{sech}^2(x_0)\right)}.
\end{equation}   
Note that $\Delta(k)$ has a zero at $k = -i/2$ as expected.\footnote{In Theorem \ref{linearizableTh} we considered only the generic case of zeros in $D_2$, whereas this zero lies on the boundary of $D_2$. However, the theory also extends to this nongeneric situation, see \cite{L-FGNLS}.} 
Moreover, note that the expression for $\tilde{\Gamma}(k)$ in (\ref{tildeGammaonesol}) coincides with the expression (\ref{Gammaonesol}) for $\Gamma(k; \infty)$ in the limit $p \to -1/2$.
This verifies that the linearizable approach gives the correct result for the stationary one-solitons $u_p^-(x, t)$.

\begin{remark}\upshape
We have established for the stationary solitons the equality
\begin{equation}\label{tildeGammalimlim}  
  \tilde{\Gamma}(k) = \lim_{p \to -1/2} \lim_{T \to \infty} \Gamma(k; T).
\end{equation}
It is important to take the limit $T \to \infty$ on the right-hand side {\it before} taking the limit $p \to -1/2$. Indeed, letting $p \to -1/2$ in (\ref{ABTonesol}) before letting $T\to \infty$, yields expressions for $A$ and $B$ different than those obtained by taking the limit $p\to -1/2$ of $A(k; \infty)$ and $B(k; \infty)$. This is due to the fact that the stationary soliton with $p = -1/2$ does {\it not} decay along the $t$-axis, so that the eigenfunction $\mu_1$ is not well-defined when $T = \infty$. If the boundary conditions had been linearizable of type (a), the equality (\ref{tildeGammalimlim}) would have followed from equation (\ref{tildeGammaGamma}). But the stationary solitons satisfy linearizable boundary conditions of type (b).

We get around this problem by first considering the soliton with $-1/2 < p<0$ for which $\mu_1$ is well-defined also for $T = \infty$, and then taking the limit $p \to -1/2$. 
It is noteworthy that, as we saw in (\ref{tildeGammaonesol}), the linearizable approach immediately gives the correct result for $\tilde{\Gamma}$. 
\end{remark}

\begin{remark}\upshape
It can be shown that the only linearizable IBV problems with initial profile given by a one-soliton $u_p^\pm(x, 0)$, $x \geq 0$, are the stationary solitons. We prove this fact for the case of an initial profile given by $u_p^-$; the case of $u_p^+$ is analogous. Let $g_0 = u_p^-(0,0)$ and $g_2 = u^-_{pxx}(0,0)$ be the boundary values of the initial data. Since $g_0 \neq 0$, linearizable boundary conditions of type (a) are ruled out. In order for a linearizable boundary condition of type (b) to exist we must have
$$g_0 = \chi, \quad g_2 = \chi + 3\chi^2, \qquad \chi \,\, \text{constant}.$$
Since $p < 0$ and
$$g_0 = -2 p^2 \text{sech}^2 x_0, \qquad
g_2 = \frac{-4 p^4 (\cosh(2 x0) -2)}{ \cosh^4(x0)},$$
this requires $p= - \frac{1}{2}$, which in turn implies that $\chi$ is given by (\ref{chionesol}).
\end{remark}

\begin{remark}\upshape
These stationary solitons may seem trivial, but actually are important for the interpretation of the asymptotics of certain solutions. Indeed, whenever a IBV problem with a linearizable boundary condition of type (b) is considered, the asymptotics of a solution must necessarily include stationary solitons in order to maintain the nonzero boundary value at $x = 0$. Equation (\ref{chionesol}) implies that for each nonzero value of $\chi > -1/2$, there are exactly two stationary solitons with value $\chi$ at $x = 0$. These two solitons are related by reflection in $x = 0$ (for $\chi = -1/2$ the two solitons coincide). For $\chi > 0$ they are of $\text{csch}^2$-type and only one of them is singularity-free on $x > 0$. For $\chi < 0$ they are of $\text{sech}^2$-type. 
Therefore, given a IBV problem with linearizable boundary condition of type (b) for some value of $\chi \geq -1/2$, we expect the asymptotics to include one of the stationary solitons with this value of $\chi$.\footnote{The nonlinear nature of the linearizable boundary condition rules out the possibility of the stationary asymptotics being accounted for by a sum of smaller stationary solitons.} It is not obvious what happens for an IBV problem with $\chi < -1/2$; perhaps this type of boundary condition will generate an infinite sequence of solitons at $x = 0$, so that a stationary state is never reached.
\end{remark}

\subsection{Two-solitons}\label{twosolitonsubsec} 
The two-solitons of equation (\ref{kdv}) and their corresponding eigenfunctions are obtained following the steps in subsection \ref{solitonsubsec} with $N=4$ and $k_3 = -\bar{k}_1$, $k_4 = -\bar{k}_2$. 
Let
$$C_1 = \frac{i e^{2 \text{x1}} p_1
   (p_1+p_2)}{p_1-p_2}, 
   \qquad 
   C_2 = \frac{i e^{2 \text{x2}} p_2
   (p_1+p_2)}{p_1-p_2},$$
   $$\theta_j = -k_j x + (k_j + 4 k_j^3)t,\qquad \Theta_j = i \theta_j + x_j, \qquad j = 1,2,$$
$k_1 = i p_1$, $k_2 = i p_2$, where the parameters satisfy\footnote{A real value of the parameter $x_j$, $j = 1,2$, corresponds to a soliton of $\text{sech}^2$-type, while if $x_j$ has imaginary part $\pi i/2$ the corresponding soliton is of $\text{csch}^2$-type.}
$$p_1 < 0, \qquad p_2 < 0, \qquad x_1 \in \R \cup \left(\R + \frac{\pi i}{2}\right), \qquad x_2 \in \RÊ\cup \left(\R + \frac{\pi i}{2}\right).$$  
Then $q$, $(\mu_3)_{12}$, $(\mu_3)_{22}$ are given by the following expressions, where we use the abbreviations $\text{ch}$ and $\text{sh}$ for $\cosh$ and $\sinh$, respectively:
\begin{equation}\label{twosoliton}
q(x,t) = -\frac{\left(p_1^2-p_2^2\right) \left(\text{ch} (2 \Theta_2)
   p_1^2+p_1^2-p_2^2+p_2^2 \text{ch} (2 \Theta_1)\right)}{(p_1 \text{ch}
   (\Theta_1) \text{ch} (\Theta_2)-p_2 \text{sh} (\Theta_1) \text{sh} (\Theta_2))^2},
\end{equation}   
\begin{align*}
(\mu_3&)_{12} = -\frac{p_1^2-p_2^2}{4 k (k+i p_1)
   (k+i p_2) (p_1 \text{ch} (\Theta_1) \text{ch} (\Theta_2)-p_2 \text{sh} (\Theta_1) \text{sh}
   (\Theta_2))^2}
   	\\
&   \times  \left(k \text{ch} (2 \Theta_2) p_1^2-i p_2
   (p_2 \text{sh} (2 \Theta_1)+p_1 \text{sh} (2 \Theta_2)) p_1+k
   \left(p_1^2-p_2^2\right)+k p_2^2 \text{ch} (2 \Theta_1)\right),
 \end{align*}  
\begin{align*}
 (\mu_3)_{22} = &
 \frac{1}{4 k (k+i p_1) (k+i p_2) (p_1 \text{ch} (\Theta_1) \text{ch}
   (\Theta_2)-p_2 \text{sh} (\Theta_1) \text{sh} (\Theta_2))^2}
   	\\
& \times \bigl[ 4 (p_1 \text{ch} (\Theta_1) \text{ch} (\Theta_2)-p_2 \text{sh} (\Theta_1) \text{sh}
   (\Theta_2))^2 k^3
   	\\
&   -4 i \left(p_1^2-p_2^2\right) \text{ch} (\Theta_2) \text{sh} (\Theta_1)
   (p_1 \text{ch} (\Theta_1) \text{ch} (\Theta_2)-p_2 \text{sh} (\Theta_1) \text{sh} (\Theta_2))
   k^2
   	\\
 &  +p_1^4k +p_2^4k
 +(p_1-p_2) (p_1+p_2) \text{ch} (2 \Theta_2) p_1^2k
   	\\
&   -p_2 \left(p_1^2+p_2^2\right) \text{sh} (2 \Theta_1) \text{sh} (2 \Theta_2)
   p_1k
   	 +p_2^2 \text{ch} (2 \Theta_1) \left(2 \text{ch} (2 \Theta_2)
   p_1^2+p_1^2-p_2^2\right)k
   	\\
 &  +i p_1 p_2
   \left(p_2^2-p_1^2\right) (p_2 \text{sh} (2 \Theta_1)+p_1 \text{sh} (2
   \Theta_2))\bigr].
\end{align*}   
We note that, as $k \to 0$,
$$\mu_3 = -\frac{i \left(p_1^2-p_2^2\right) (p_2 \sinh (2 \Theta_1)+p_1 \sinh (2
   \Theta_2))}{4 k (p_1 \cosh (\Theta_1) \cosh (\Theta_2)-p_2 \sinh (\Theta_1) \sinh
   (\Theta_2))^2} \begin{pmatrix} 1 & 1 \\ -1 & -1 \end{pmatrix} + O(1),$$
in accordance with (\ref{muatorigin}).

\subsubsection{Spectral functions}
For simplicity, we now assume that $x_1 = x_2 = 0$.
Evaluating $\mu_3$ at $x = t = 0$, we find the spectral functions
$$a(k) =\frac{2 k^2+p_1^2+p_2^2}{2 (k+i p_1) (k+i p_2)}, \qquad b(k) = \frac{p_2^2-p_1^2}{2 (k+i p_1) (k+i p_2)}.$$  
The spectral functions for the problem on the line are given by
$$a^L(k)= \frac{(k-i p_1) (k-i p_2)}{(k+i p_1) (k+i p_2)}, \qquad b^L(k) = 0.$$
The zeros of $a^L(k)$, as expected, occur at $k = ip_1$ and $k = ip_2$.

\begin{figure}
\begin{center}
\begin{tabular}{cc}
Profile at $t = 50$ & $|d(k; \infty)|$ 
	\\ \hline
     \includegraphics[width=.3\textwidth]{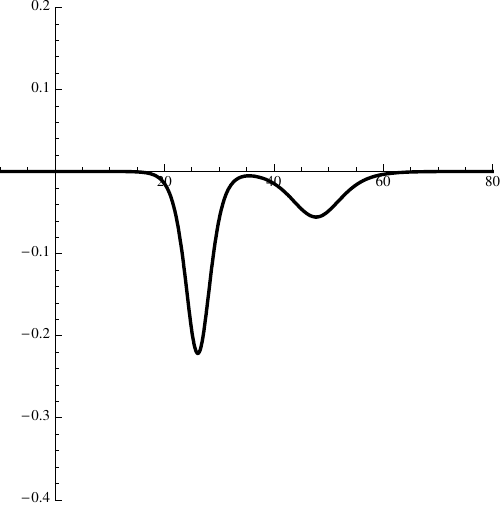} 
     &
   \includegraphics[width=.3\textwidth]{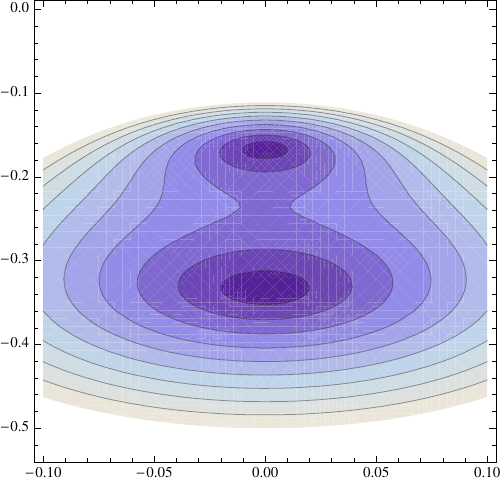} \quad
   \end{tabular}
      \begin{figuretext}\label{twosol.pdf} \upshape
         The two-soliton with $x_1 = x_2 = 0$, $p_1 = -1/3$ and $p_2 = -1/6$ at time $t = 50$ (left) and a contour plot of the function $|d(k; \infty)|$ whose zeros generate the asymptotic solitons (right). A dark (light) area indicates a small (large) value of $|d(k; \infty)|$; the $x$ and $y$ axes are labeled by $\text{Re}\,k \in(-0.1, 0.1)$ and $\text{Im}\,k \in (-0.54, 0)$, respectively. The zeros lie at $k = -i/3$ and $k = -i/6$ as expected.
       \end{figuretext}
   \end{center}
\end{figure}

The expressions for $\mu_2$, $S(k)$, and $d(k)$ are too long to present in general, but we can illustrate the results in some special cases. 
Let $p_1 = -1/3$ and $p_2 = -1/6$. Then
$$d(k; T) = \frac{1} {2 k \left(18 k^2-9 i k-1\right) \left(3 \cosh\left(\frac{T}{27}\right)+\cosh \left(\frac{T}{3}\right)\right)^2} 
   \biggl[108 \cosh \left(\frac{10 T}{27}\right) k^3$$
 $$+18 \cosh \left(\frac{2 T}{3}\right) k^3+180 k^3+27 i
   \sinh \left(\frac{2 T}{27}\right) k^2 +18 i \sinh \left(\frac{8 T}{27}\right) k^2$$
   $$+36 i \sinh
   \left(\frac{10 T}{27}\right) k^2 +9 i \sinh \left(\frac{2 T}{3}\right) k^2+3 \cosh \left(\frac{10
   T}{27}\right) k-\cosh \left(\frac{2 T}{3}\right) k+17 k$$
   $$+9 \left(18 k^3+k\right) \cosh \left(\frac{2
   T}{27}\right)+12 \left(9 k^3+k\right) \cosh \left(\frac{8 T}{27}\right)+2 i \sinh \left(\frac{8
   T}{27}\right) +i \sinh \left(\frac{10 T}{27}\right)\biggr].$$
As $T \to \infty$, we find, for $k \in \bar{D}_2$,
$$d(k; \infty) = \frac{-18 k^2-9 i k+1}{-18 k^2+9 i k+1},$$
with zeros at $k = ip_1 = -i/3$ and $k = ip_2 = -i/6$ as expected, see Figure \ref{twosol.pdf}. Also, in the limit $T \to \infty$, we have
$$A(k; \infty) = \frac{72 k^2+5}{72 k^2+36 i k-4}, \qquad B(k; \infty) = \frac{3}{-72 k^2-36 i k+4},$$
and we can easily verify the global relation
$$B(k; \infty) a(k) - A(k; \infty) b(k) = 0, \qquad k \in \bar{D}_1.$$
Furthermore, the expressions for the spectral functions have the appropriate behavior as $k \to \infty$ and $k \to 0$.

\subsubsection{Linearizable two-soliton}
\begin{figure}
\begin{center}
\begin{tabular}{cc}
Profile at $t = 50$ & $|\Delta(k)|$ 
	\\ \hline
     \includegraphics[width=.3\textwidth]{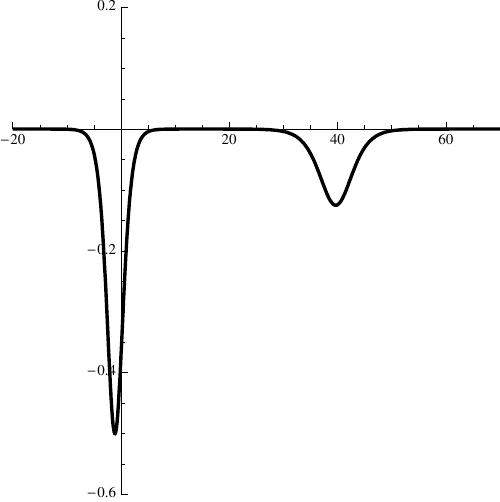} 
     &
   \includegraphics[width=.3\textwidth]{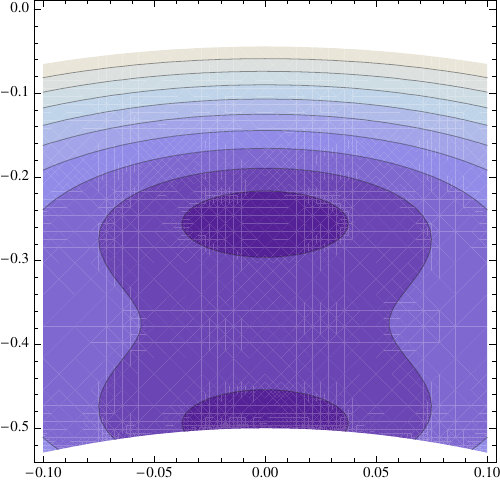} \quad
   \end{tabular}
      \begin{figuretext}\label{lintwosol.pdf} \upshape
         The linearizable two-soliton with $x_1 = x_2 = 0$, $p_1 = -1/2$ and $p_2 = -1/4$ at time $t = 50$ (left) and a contour plot of the function $|\Delta(k)|$ whose zeros generate the asymptotic solitons (right). The zeros lie at $k = -i/2$ and $k = -i/4$ as expected.
       \end{figuretext}
   \end{center}
\end{figure}

Retaining the assumption $x_1 = x_2 = 0$ and letting
$$p_1 = -1/2, \quad p_2 = -1/4,$$
we obtain the solution
$$u(x,t) = -\frac{3 \left(4 \cosh \left(\frac{1}{8} (3 t-4 x)\right)+\cosh
   (x)+3\right)}{4 \left(\cosh \left(\frac{3}{16} (t-4 x)\right)+3 \cosh
   \left(\frac{1}{16} (3 t+4 x)\right)\right)^2},$$
which satisfies a linearizable boundary condition of type (b) with $\chi = -\frac{3}{8}$.
From the expressions for the spectral functions,
$$\begin{pmatrix} b(k) \\ a(k) \end{pmatrix} = \left(
\begin{array}{l}
 \frac{3}{-32 k^2+24 i k+4} 
 \\
 -\frac{32 k^2+5}{-32 k^2+24 i k+4}
\end{array}
\right),$$
we obtain $\Delta(k)$ according to (\ref{bDeltadef}). The result is
$$\Delta(k) = -\frac{4}{3} \left(8 k^2+6 i k-1\right).$$
As expected, $\Delta(k)$ has two zeros at $k = -i/2$ and $k = -i/4$, see Figure \ref{lintwosol.pdf}.

\subsection{Rational solitons}
In addition to the solitons with exponential decay, equation (\ref{kdv}) also admits solitons with algebraic decay at infinity. 
Taking the limit $p \to 0$ in the formula (\ref{sinhonesoliton}) for the $\text{csch}^2$-soliton with $x_0 = 0$, we find the rational solution
\begin{equation}\label{rationalsol}  
  u(x,t) = \frac{2}{(x - x_0 -t)^2},
\end{equation}
where we have introduced a new parameter $x_0$ in the answer to allow a translational shift of the initial data.
The corresponding eigenfunction can be found by taking the same limit in (\ref{onesolitonmu3}). 
Similarly, letting $x_1 = x_2 =\frac{i \pi }{2}$ in the formula (\ref{twosoliton}) for the two-soliton and taking the limit $p_1,p_2 \to 0$, we find the rational two-soliton
$$u(x,t) = \frac{6 (t-x) \left(t^3-3 x t^2+3 \left(x^2-8\right)
   t-x^3\right)}{\left(t^3-3 x t^2+3 \left(x^2+4\right) t-x^3\right)^2}.$$
Although the rational solitons have singularities when considered on the line, for the case of the half-line it is always possible to choose these singularities to lie outside the half-line, at least initially. It is therefore natural to consider examples of IBV problems with initial data given by the profile of some rational soliton. However, these solutions do not have sufficient decay at $x = \infty$ for the formalism of section \ref{spectralsec} to be applicable.
For example, the spectral functions $a(k)$ and $b(k)$ for an initial profile given by the rational one-soliton (\ref{rationalsol}) at time $t = 0$ with $x_0 < 0$ is given by
$$\begin{pmatrix} b(k) \\ a(k) \end{pmatrix} 
= \left(
\begin{array}{l}
 \frac{1}{2 k^2 x_0^2} \\
 1+\frac{i}{k x_0}-\frac{1}{2 k^2 x_0^2}
\end{array}
\right).$$
It is clear that these spectral functions do {\it not} have the correct behavior (\ref{abkto0}) as $k \to 0$. The study of rational solitons lies beyond the spectral theory defined in section \ref{spectralsec} and therefore we will not consider these examples any further.

\section{Exponential initial profiles}\label{exponentialsec} \nequation
Our next class of examples consists of IBV problems whose initial data is given by an exponential function.
The initial and boundary conditions (\ref{Erxdefintro})-(\ref{expIBVintro}) describe an IBV problem for equation (\ref{kdv}). Since the boundary conditions are of the form (b) of Theorem \ref{linearizableTh}, they are linearizable. Therefore, if we can find the spectral functions $a(k)$ and $b(k)$ from the initial data, then we can obtain the function $\Delta(k)$ from equation (\ref{bDeltadef}) and the zeros of $\Delta(k)$ will give us information about the asymptotic behavior of the solution $u_r(x,t)$ for large $t$.
Observe that the initial data $E_r(x)$ of (\ref{Erxdefintro}) grows exponentially as $x \to -\infty$; hence there is no obvious extension of this problem to the line.

\begin{remark}\upshape
The particular form of the initial profiles $E_r(x)$ can be motivated as follows: The simplest possible exponential profiles with decay as $x \to \infty$ are of the form $u_0(x) = c e^{-r x + \alpha}$ for some values of $c, \alpha \in \R$, and $r > 0$. Letting $\chi := u_0(0) = c e^{\alpha}$, the condition that this profile is compatible with a linearizable boundary condition at the point $x = t = 0$, takes the form
$$u_{0xx}(0) = \chi + 3 \chi^2, \qquad \text{i.e.} \qquad  c e^\alpha r^2 = c e^{\alpha} + 3 c^2 e^{2 \alpha}.$$
Solving this equation for $c$ and discarding the trivial solution $c = 0$ yields
$$c = \frac{1}{3} e^{-\alpha } \left(r^2-1\right), \qquad \text{i.e.} \qquad u_0(x) = E_r(x).$$
Hence, the profiles $E_r(x)$ defined in (\ref{Erxdefintro}) are the simplest exponential initial profiles compatible with a linearizable IBV problem.
\end{remark}

\subsection{Determination of $a(k)$ and $b(k)$}
In order to find analytic expressions for the spectral functions $a(k)$ and $b(k)$, we will analyze the $x$-part of the scalar Lax pair for (\ref{kdv}) presented in Appendix \ref{scalarlaxapp}. In particular,
\begin{equation}\label{scalarxpart}
  \varphi_{xx} = (u + k^2)\varphi, \qquad u(x) = E_r(x).
\end{equation}
We will then use equation (\ref{varphitomu3}) to determine the matrix eigenfunction $\mu_3(x,0,k)$, from which $a(k)$ and $b(k)$ are defined according to 
$$\begin{pmatrix} b(k) \\ a(k) \end{pmatrix} = [\mu_3(0,0,k)]_2.$$ 
The general solution of equation (\ref{scalarxpart}) depends on two constants $c_1, c_2$ and is given by 
\begin{align} \label{varphigeneral}
\varphi(x,k; c_1, c_2) = & (-1)^{-\frac{k}{r}} I_{-\frac{2 k}{r}}\left(\frac{2 \sqrt{e^{-r x}}
   \sqrt{r^2-1}}{\sqrt{3} r}\right) c_1 \Gamma \left(1-\frac{2
   k}{r}\right)
  	\\ \nonumber
&  +(-1)^{k/r} I_{\frac{2 k}{r}}\left(\frac{2 \sqrt{e^{-r x}}
   \sqrt{r^2-1}}{\sqrt{3} r}\right) c_2 \Gamma \left(\frac{2
   k}{r}+1\right),
\end{align}
where $I_a(x)$ denotes the modified Bessel function of the first kind and $\Gamma(z)$ is the Euler gamma function. We denote by $\varphi(x,k)$ the particular eigenfunction with the asymptotic behavior $\varphi(x,k) \sim e^{kx}$ as $x\to \infty$. This eigenfunction can be found from (\ref{varphigeneral}) by an appropriate choice of $c_1$ and $c_2$. Since the Bessel function $I_a(x)$ has the asymptotic behavior
   $$I_a(x) \sim x^a \left(\frac{2^{-a}}{\Gamma (a+1)}+O\left(x^2\right)\right), \qquad x \to 0,$$
it follows that $\varphi(x, k) = \varphi(x,k; c_1, c_2)$, where
$$c_1 = \left(-\frac{1}{3}\right)^{k/r}
   \left(\frac{\sqrt{r^2-1}}{r}\right)^{\frac{2 k}{r}}, \qquad c_2 = 0,$$
i.e.
$$\varphi(x,k) = 3^{-\frac{k}{r}} \left(\frac{\sqrt{r^2-1}}{r}\right)^{\frac{2 k}{r}}
   I_{-\frac{2 k}{r}}\left(\frac{2 \sqrt{e^{-r x}} \sqrt{r^2-1}}{\sqrt{3}
   r}\right) \Gamma \left(1-\frac{2 k}{r}\right).$$
The matrix eigenfunction $\mu_3(x,0,k)$ is obtained from $\varphi(x,k)$ via equation (\ref{varphitomu3}). This yields
$$[\mu_3(x,0,k)]_2 = \begin{pmatrix}
\frac{i e^{-r x} \left(r^2-1\right) \, _0F_1\left(;\frac{2 i
   k}{r}+2;\frac{e^{-r x} \left(r^2-1\right)}{3 r^2}\right)}{6 k (2 i
   k+r)}
 	\\  
  \, _0F_1\left(;\frac{2 i k}{r};\frac{e^{-r x} \left(r^2-1\right)}{3
   r^2}\right)
   \end{pmatrix},$$
where $\, _0F_1(; \cdot; \cdot)$ is the confluent hypergeometric function cf. \cite{Slater}. It can be explicitly verified that the expression for $\mu_3(x,0,k)$ has the correct asymptotic behavior, $\mu_3(x,0,k) \to I$ as $x \to \infty$, and that is satisfies the $x$-part of the Lax pair (\ref{lax}) with potential $u = E_r(x)$.
Evaluating $\mu_3(x,0,k)$ at $x = 0$ yields
\begin{align} \label{expab}
&b(k) =-\frac{\left(r^2-1\right) \, _0F_1\left(;\frac{2 i
   k}{r}+2;\frac{1}{3}-\frac{1}{3 r^2}\right)}{6 k (i r-2 k)},
   \\ \nonumber
&a(k) = \, _0F_1\left(;\frac{2 i k}{r};\frac{1}{3}-\frac{1}{3 r^2}\right).
\end{align}
Note that $\, _0F_1(;i k;z)$ is analytic as a function of $k$ except for a countable number of simple poles at $k = i n$, $n = 0,1,2, \dots$. The modified Bessel function $I_{\nu }(z)$ is analytic in the whole complex plane as a function of $\nu$. In particular, $a(k)$ and $b(k)$ are analytic for $k \in D_1 \cup D_2$ as required.
It can be directly verified, in accordance with (\ref{abkto0}), that
$$a(k) = \frac{i \alpha}{k} + O(1), \quad b(k) = -\frac{i\alpha}{k} + O(1), \qquad k \to 0,$$
where 
$$\alpha = -\frac{\sqrt{r^2-1} I_1\left(\frac{2 \sqrt{r^2-1}}{\sqrt{3} r}\right)}{2
   \sqrt{3}}.$$

\subsection{Asymptotic behavior}
Substituting the expressions for $a(k)$ and $b(k)$ given in (\ref{expab}) into (\ref{bDeltadef}) yields an explicit formula for $\Delta(k)$. The function $\Delta(k)$ is defined for $k \in \bar{D}_2$ and its zeros lie on the imaginary axis. We denote these zeros by $\{ip_n\}_1^N$, $-1/2 \leq p_1 < p_2 < \cdots < p_N < 0$. Numerical computations suggest that the zeros are simple.
As described in subsection \ref{asymptoticbehaviorsubsec}, the solution $u_r(x,t)$ for large $t$ splits into a sum of one-solitons, $u_r(x,t)= \sum_{n = 1}^N u_{p_n}(x,t) + o(1)$. For $-1/2 < p_n < 0$, the soliton $u_{p_n}(x,t)$ must be of the form (\ref{onesoliton}),
$$u_{p_n}(x,t) = -2p_n^2 \text{sech}^2\left(4 t p_n^3+(x-t) p_n + x_{0n}\right),$$
while if $p_1 = -1/2$ it can also be of the form (\ref{sinhonesoliton}).

\begin{figure}
\begin{center}
\begin{tabular}{ccc}
$r$ & \begin{tabular}{ll}  Initial profile $E_r(x)$ \qquad \qquad& \qquad \quad$|\Delta(k)|$\qquad \quad \end{tabular} & $\sharp$ of solitons
	\\ \hline
	$r = 5$ & \begin{tabular}{cc}
     \includegraphics[width=.3\textwidth]{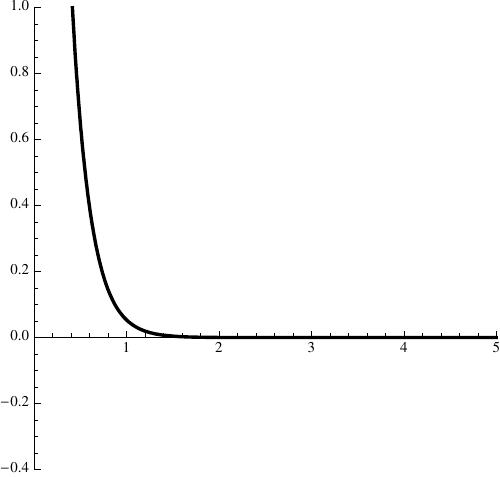} \quad
   \includegraphics[width=.3\textwidth]{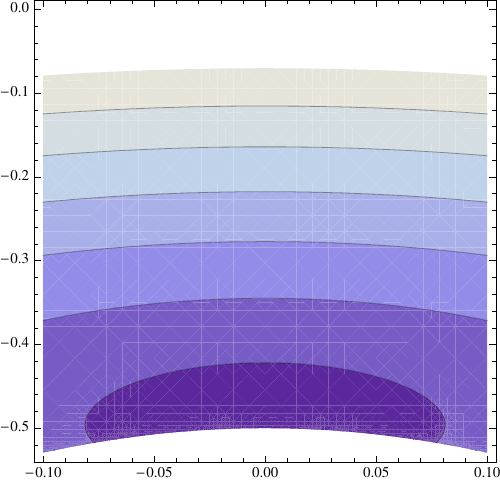} \quad
   \end{tabular}
   & $\begin{array}{c} 1 \\ _{\begin{pmatrix} p_1 = -1/2 \end{pmatrix}} \end{array}$
   	\\
	$r = 2.8$ & \begin{tabular}{cc}
   \includegraphics[width=.3\textwidth]{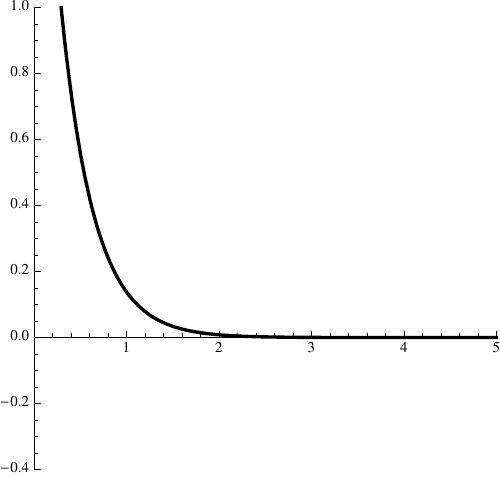} \quad
   \includegraphics[width=.3\textwidth]{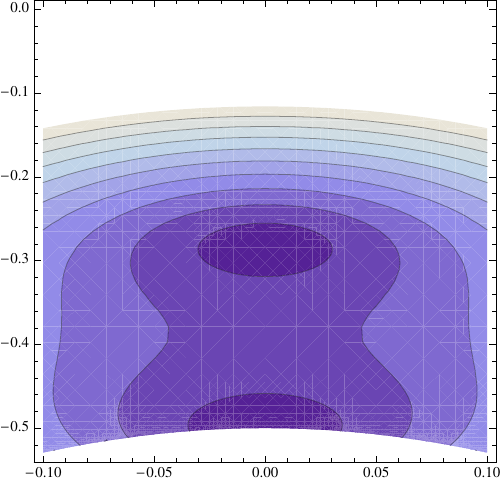} \quad
     \end{tabular}
    & $\begin{array}{c} 2 \\ _{\begin{pmatrix} p_1 = -1/2 \\ p_2 = -0.28 \end{pmatrix}} \end{array}$
	\\
	$r = 2$ & \begin{tabular}{cc}
     \includegraphics[width=.3\textwidth]{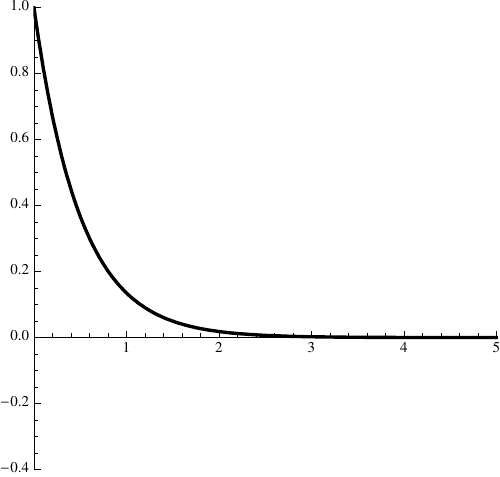} \quad
   \includegraphics[width=.3\textwidth]{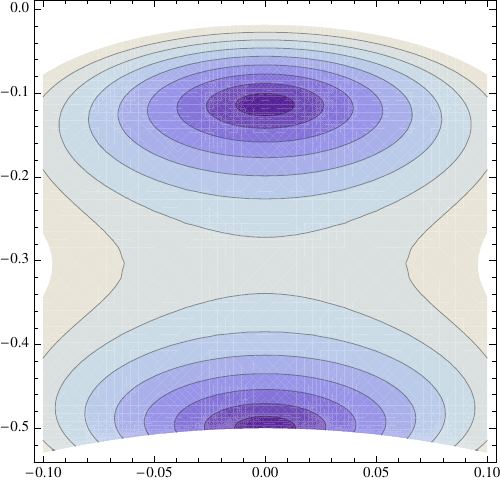} \quad
   \end{tabular}
   & $\begin{array}{c} 2 \\ _{\begin{pmatrix} p_1 = -1/2 \\ p_2 = -0.11 \end{pmatrix}} \end{array}$
      	\\
	$r = 1.1$ & \begin{tabular}{cc}
    \includegraphics[width=.3\textwidth]{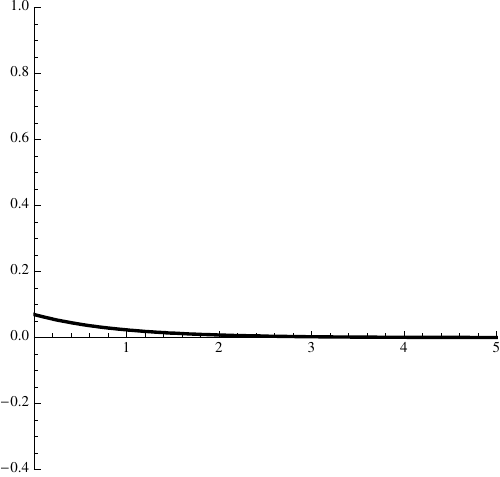} \quad
   \includegraphics[width=.3\textwidth]{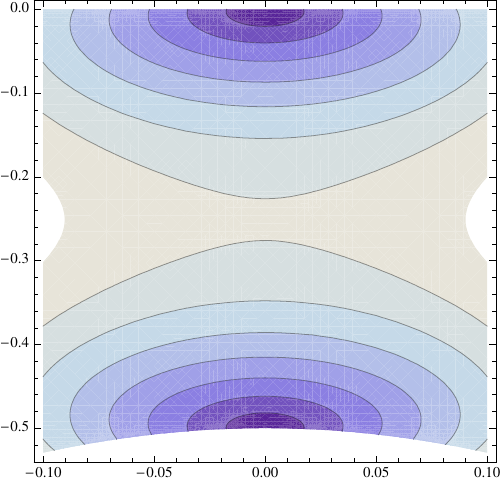} \quad
   \end{tabular}
   & $\begin{array}{c} 2 \\ _{\begin{pmatrix} p_1 = -1/2 \\ p_2 = -0.0020  \end{pmatrix}}\end{array}$
   	\\
\end{tabular}
     \begin{figuretext}\label{expfigure} \upshape
              For the values of $r$ specified in the first column, the second column shows the initial profile $u_r(x,0) = E_r(x)$ for the linearizable initial-boundary value problem (\ref{expIBVintro}) (the $x$ and $y$ axes are labeled by $x$ and $E_r(x)$, respectively). The third column displays contour plots of $|\Delta(k)|$ (a dark (light) area indicates a small (large) value of $|\Delta(k)|$; the $x$ and $y$ axes are labeled by $\text{Re}\,k \in(-0.1, 0.1)$ and $\text{Im}\,k \in (-0.54, 0)$, respectively). The fourth column shows the number of solitons exhibited asymptotically for large $t$ by the solution $u_r(x,t)$ and the corresponding values of $p$. The solitons correspond to the zeros of $\Delta(k)$.     \end{figuretext}
     \end{center}
\end{figure}

The associated expression for $\Delta(k)$ is rather complicated and not suitable for analytical methods. However, numerical methods suggest the following picture, see Figure \ref{expfigure}: 
\begin{itemize}

\item For any value of $r$, $\Delta(k)$ has a zero at $k = -i/2$. 

\item For $r >> 1$, the only zero of $\Delta(k)$ is $k = -i/2$. 

\item For $0.25 \lesssim r \lesssim  3.33$ and $r \neq 1$, $\Delta(k)$, in addition to the zero at $k = -i/2$, has {\it one} more zero.
This zero enters $D_2$ from below at $k = -i/2$ as $r$ decreases below $3.33$; as $r$ decreases further the zero moves upward along the imaginary axis; for $r \to 1$ it hits the origin $k = 0$ and turns around; as $r$ decreases further the zero moves downward along the imaginary axis.

\item For $0.15 \lesssim r \lesssim 0.25$, $\Delta(k)$, in addition to the zero at $k = -i/2$, has {\it two} more zeros. The new zero enters $D_2$ from above at $k = 0$ as $r$ decreases below $0.25$; as $r$ decreases below $0.15$ yet another zero enters $D_2$ from above.

\item As $r \downarrow 0$, the number of zeros of $\Delta(k)$ increases. New zeros keep entering $D_2$ from above.
For $r \approx 0.05$, in addition to the zero at $k = -i/2$, there exist {\it eight} more zeros.
\end{itemize}

\begin{figure}
\begin{center}
\begin{tabular}{ccc}
$r$ &  \begin{tabular}{ll}  Initial profile $E_r(x)$ \qquad \qquad& \qquad \quad$|\Delta(k)|$\qquad \quad \end{tabular}  & $\sharp$ of solitons
	\\ \hline
	$r = 0.9$ & \begin{tabular}{cc}
    \includegraphics[width=.3\textwidth]{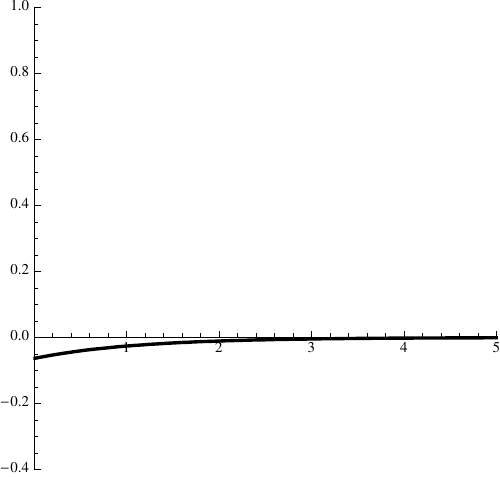} &
   \includegraphics[width=.3\textwidth]{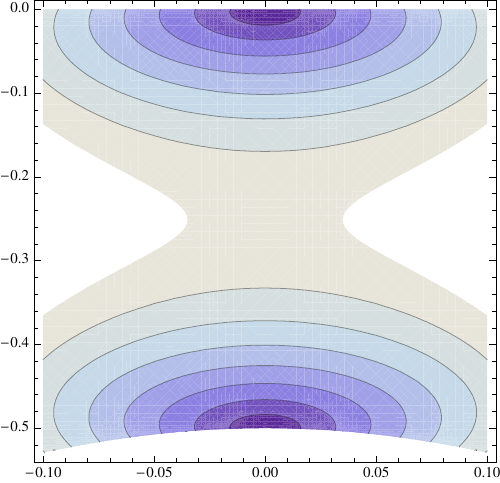}
   \end{tabular}
     & $\begin{array}{c} 2 \\ _{\begin{pmatrix} p_1 = -1/2 \\ p_2 = -0.0024 \end{pmatrix}} \end{array}$
        	\\
	$r = 0.5$ & \begin{tabular}{cc}
    \includegraphics[width=.3\textwidth]{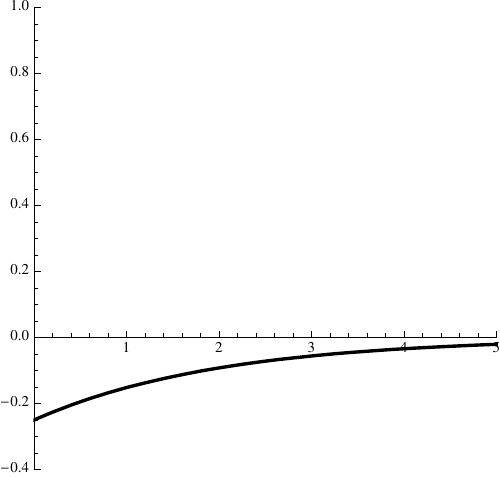} &
   \includegraphics[width=.3\textwidth]{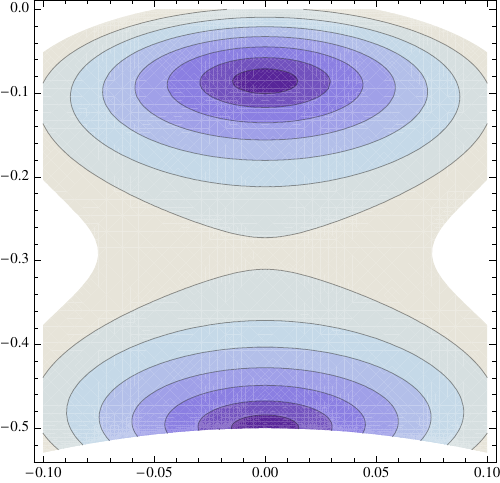}
     \end{tabular}
    & $\begin{array}{c} 2 \\ _{\begin{pmatrix} p_1 = -1/2 \\ p_2 = -0.085 \end{pmatrix}} \end{array}$
	\\
	$r = 0.2$ & \begin{tabular}{cc}
    \includegraphics[width=.3\textwidth]{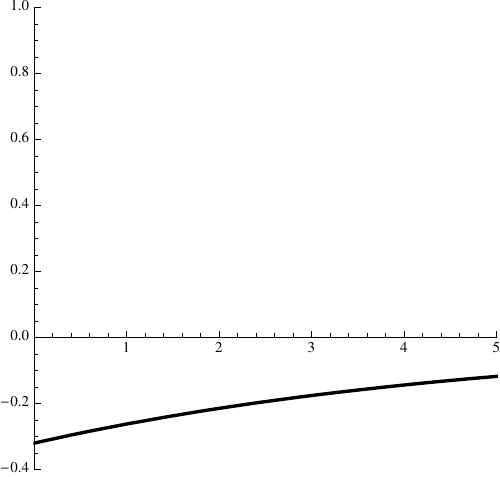} &
   \includegraphics[width=.3\textwidth]{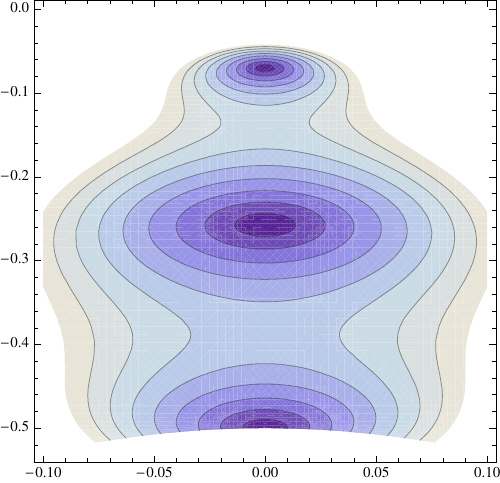}
   \end{tabular}
   & $\begin{array}{c} 3 \\ _{\begin{pmatrix} p_1 = -1/2  \\ p_2 = -0.26 \\ p_3 = -0.070 \end{pmatrix}} \end{array}$
   	\\
	$r = 0.14$ & \begin{tabular}{cc}
    \includegraphics[width=.3\textwidth]{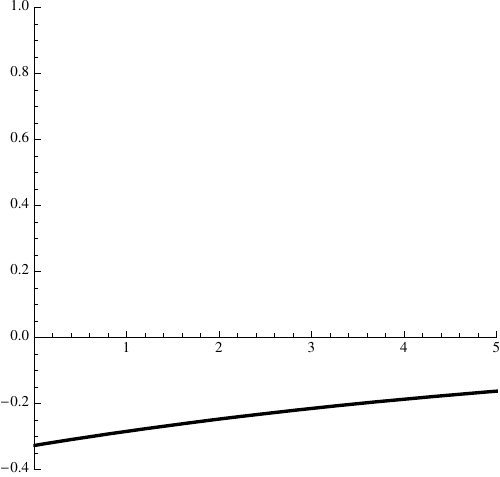} &
   \includegraphics[width=.3\textwidth]{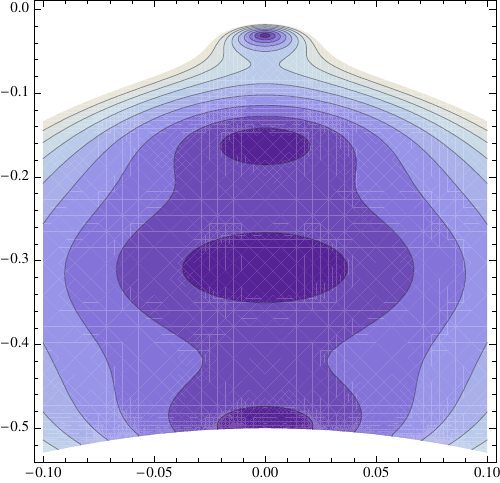}
   \end{tabular}
    & $\begin{array}{c} 4 \\ _{\begin{pmatrix} p_1 = -1/2 \\ p_2 = -0.31 \\ p_3 = -0.16 \\ p_4 = -0.031 \end{pmatrix}} \end{array}$
  	\\
\end{tabular}
          \end{center}
{\bf Figure \ref{expfigure} (continued).}
\end{figure}

\begin{figure}
\begin{center}
    \includegraphics[width=.6\textwidth]{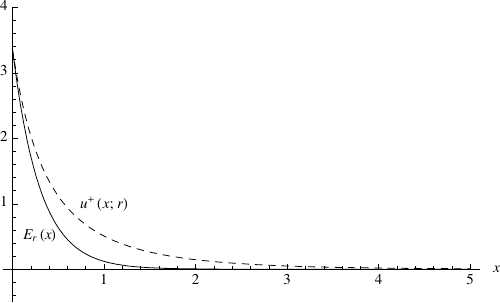}
     \begin{figuretext}\label{Erxanduplus.pdf}
        The initial profile $E_r(x)$ (solid) and the stationary soliton $u^+(x; r)$ (dashed) for $r = 3.33$.
     \end{figuretext}
     \end{center}
\end{figure}

\begin{figure}
\begin{center}
    \includegraphics[width=.6\textwidth]{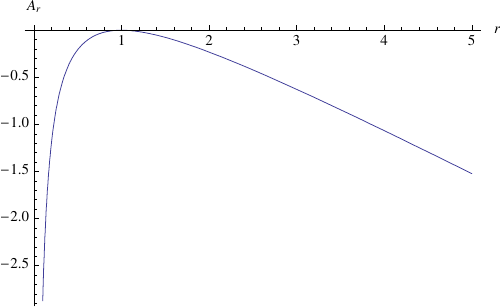}
     \begin{figuretext}\label{areadiff.pdf}
        The area difference $A_r = \int_0^\infty (E_r(x) - u^+(x; r))dx$ plotted as a function of $r$.
     \end{figuretext}
     \end{center}
\end{figure}

\begin{remark}\upshape
These observations regarding the zeros of $\Delta(k)$ can be interpreted in terms of asymptotic solitons as follows. The zero at $k = -i/2$, which is present for all $r$, generates a stationary soliton. As discussed in subsection \ref{linonesolsubsubsec}, this soliton is needed in order to maintain the nonzero boundary value at $x = 0$. For $r > 1$, we have $\chi_r > 0$, so that this soliton is of $\text{csch}^2$-type. For $0 < r < 1$, we have $-1/3 < \chi_r < 0$, so that it is of $\text{sech}^2$-type.

The reason for the appearance of another zero as $r$ decreases below $3.33$ is not clear, but the following seems a plausible explanation. First assume $r > 1$. Near the boundary, we expect the solution $u_r(x,t)$ to attempt to evolve into the stationary soliton of $\text{csch}^2$-type with value $E_r(0) = \chi_r$ at $x = 0$. This soliton has data $p = -1/2$ and $x_0 = - \text{csch}^{-1}\left(\sqrt{\frac{2}{3}} \sqrt{r^2-1}\right)$; we denote it by $u^+(x; r)$.
Comparing the profiles $E_r(x)$ and $u^+(x; r)$, we see that $E_r(x)$ lies below $u^+(x; r)$, see Figure \ref{Erxanduplus.pdf}. If $u(x,t)$ were to evolve from the initial profile $E_r(x)$ into the stationary profile $u^+(x; r)$, then the solution would have to increase near $x = 0$. In view of the Remark \ref{conservedremark}, an increase in $u(x,t)$ near $x = 0$ must be accompanied by a decrease of $u(x,t)$ for larger values of $x$. Considering that the shape of the difference $E_r(x) - u^+(x; r)$ is quite similar to that of a $\text{sech}^2$-soliton, it is natural to expect that in the process of raising the solution from $E_r(x)$ to $u^+(x; r)$ near the boundary, another one-soliton $u_p^-(x,t)$ is sent off toward $x \to \infty$. However, since the only right-traveling solitons are the ones with $-1/2 < p < 0$, only solitons within this range can be generated. For $r >> 1$ the difference in area between $E_r(x)$ and $u^+(x; r)$ is very large, while as $r \downarrow 1$ it shrinks to zero, see Figure \ref{areadiff.pdf}. It is therefore possible for a right-traveling soliton to appear asymptotically whenever $r$ falls below the value for which the size of the difference $E_r(x) - u^+(x; r)$ equals that of the largest right-traveling soliton. As $r$ decreases even further, we expect the size of the soliton $u_p^-(x,t)$ to decrease at a rate comparable to the rate at which $E_r(x) - u^+(x; r)$ decreases. In fact, since we expect the $\text{sech}^2$-soliton to be accompanied by some dispersive wave trains of predominantly positive elevation, it is plausible that the size of the generated soliton is actually larger than the value suggested by the area difference
$$A_r = \int_0^\infty (E_r(x) - u^+(x; r))dx.$$
For example, for the box example in the following section we will find that an initial profile with $\int_0^\infty u(x,0) dx = -1$, generates an asymptotic soliton with area $-1.7404$.
In order to test this interpretation we compute
\begin{equation}\label{onesolarea}  
  \int_{-\infty}^\infty u_p^-(x, t) dx = 4 p
\end{equation}
and
$$\int_0^\infty E_r(x) dx = \frac{r^2-1}{3 r}, \qquad \int_0^\infty u^+(x; r) dx = \frac{\sqrt{2 r^2+1}}{\sqrt{3}}-1.$$
Hence,
\begin{equation}\label{Ardef}
  A_r = \frac{r \left(r-\sqrt{6 r^2+3}+3\right)-1}{3 r}.
\end{equation}
Let $p_r$ denote the value of $p$ for the asymptotic soliton generated by $E_r(x)$ for $1 < r \lesssim 3.33$.
The first soliton appears when $r \approx 3.33$ and has area $4\cdot p_{3.33} = -2$.
On the other hand, the area difference for $r \approx 3.33$ is $A_r \approx -0.77$.
In Table \ref{areatable} we compare the area difference $A_r$ with the corresponding size of the generated soliton for some values of $r$. 
In all cases the area of the generated soliton is about twice that of $A_r$. This is in good agreement with the observations for the box profile already mentioned.
\end{remark}

\begin{table}[htdp]
\caption{Comparing $A_r$ with the size of the generated soliton.}\label{areatable}
\begin{center}
\begin{tabular}{|c|c|c|c|}
$r $ & $A_r$ & $p_r$ & $\int_{-\infty}^\infty u_{p_r}^-(x, t) dx$ \\ \hline
3.33 & -0.77 &  -0.5 & -2 \\
2.8 & -0.77 &  -0.28 & -1.12 \\
2   & -0.23  & -0.11 & -0.44 \\
1.1  & -0.0041 & -0.002 & -0.0080 \\
\end{tabular}
\end{center}
\label{default}
\end{table}%

\begin{remark}\upshape
The interpretation of the observations for $0 < r < 1$ is similar, except that the stationary soliton now is of $\text{sech}^2$-type since $E_r(0) < 0$. Denoting this stationary soliton by $u^-(x; r)$ we find that the area difference $A_r = \int_0^\infty (E_r(x) - u^-(x; r))dx$ is given by formula (\ref{Ardef}) just as before. However, this time when the area difference increases beyond that of the largest right-traveling soliton, 
a collection of small solitons is generated, since no single soliton can compensate for this difference.
\end{remark}

\section{Box-shaped initial profiles}\nequation\label{boxsec}
In this section we consider the IBV problem for equation (\ref{kdv}) with initial and boundary conditions given by (\ref{boxIBVintro}).
The initial profile $\beta_I(x)$ is a box of height $h$ and length $L$ positioned with its left edge at $x = x_0$.
The boundary conditions are linearizable of type (a) and we can proceed in the same way as in the previous section: (1) We find the eigenfunction $\mu_3(x,0,k)$ by solving the $x$-part of the Lax pair with the potential given by $\beta_I(x)$. (2) We find the spectral functions $a(k)$ and $b(k)$ by evaluating $\mu_3(x,0,k)$ at $x = 0$. (3) We find $\Delta(k)$ from equation (\ref{aDeltadef}). (4) We investigate the asymptotic behavior of $u_I(x,t)$ by analyzing numerically the zeros of $\Delta(k)$.

\subsection{Determine $a(k)$ and $b(k)$}
Consider the matrix equation
\begin{equation}\label{psixpartbox}  
  \psi_x - ik\sigma_3 \psi = V_1 \psi,
\end{equation}
where $\psi$ is a $2 \times 2$ matrix valued function and $V_1$ is defined in terms of the initial data $u_I(x,0)$ by equation (\ref{V1V2def}). If we can find a solution $\psi(x,k)$ of this equation with the asymptotic behavior $\psi(x,k) \sim e^{ik x \sigma_3}$, then the eigenfunction $\mu_3(x,0,k)$ is related to $\psi(x,k)$ by 
\begin{equation}\label{psimu3box}  
  \psi(x,k) = \mu_3(x,0,k) e^{ik x \sigma_3}.
\end{equation}
For $x > x_0 + L$, equation (\ref{psixpartbox}) is simply
\begin{equation}\label{psixpartsimple}  
  \psi_x - ik\sigma_3 \psi = 0,
\end{equation}
so that
$$\psi(x,k) = e^{ik x \sigma_3}, \qquad x > x_0 + L.$$
Letting
$$[\psi(x,k)]_2 = \begin{pmatrix} \psi_1 \\ \psi_2 \end{pmatrix},$$
the second column of equation (\ref{psixpartbox}) for $x \in (x_0, x_0 + L)$ reads
\begin{align*} 
\left(h-2 k^2\right) \psi_1 +h \psi_2-2 i k \psi_{1x} = 0&, 
	\\
  \left(h-2 k^2\right) \psi_2 + h \psi_1+2 i k \psi_{2x} = 0&. \end{align*}
We find
\begin{align*}
\psi_1(x,k) =& e^{-\frac{\alpha x}{k}} \left(2 i e^{\frac{2 \alpha
   x}{k}} k^2-2 i k^2-i e^{\frac{2 \alpha x}{k}} h+i h+2 e^{\frac{2
   \alpha x}{k}} \alpha+2 \alpha\right) \frac{c_1}{4 \alpha}
   	\\
&   -i e^{-\frac{\alpha x}{k}} \left(-1+e^{\frac{2
   \alpha x}{k}}\right) \frac{h c_2}{4 \alpha},
   	\\
 \psi_2 (x,k) =  & e^{-\frac{\alpha
   x}{k}} \left(-2 i e^{\frac{2 \alpha x}{k}} k^2+2 i k^2+i
   e^{\frac{2 \alpha x}{k}} h-i h+2 e^{\frac{2 \alpha x}{k}}
   \alpha+2 \alpha\right) \frac{c_2}{4 \alpha}
   	\\
  &+ i e^{-\frac{\alpha x}{k}} \left(-1+e^{\frac{2 \alpha
   x}{k}}\right) \frac{h c_1}{4 \alpha},
 \end{align*}
where $\alpha = \sqrt{h k^2-k^4}$ and $c_1, c_2$ are two parameters which need to be determined so that $\psi(x,k)$ is continuous at $x = x_0 + L$. A straightforward computation yields
\begin{align*}
&c_1 = -\frac{i e^{-\frac{\left(i k^2+\alpha\right) (L+x_0)}{k}}
   \left(-1+e^{\frac{2 \alpha (L+x_0)}{k}}\right) h
   \alpha}{4 k^2 \left(k^2-h\right)},
 	\\  
& c_2 = -\frac{i e^{-\frac{\left(i k^2+\alpha\right) (L+x_0)}{k}}
   \left(2 \left(-1+e^{\frac{2 \alpha
   (L+x_0)}{k}}\right) k^2-e^{\frac{2 \alpha
   (L+x_0)}{k}} h+h-2 i \left(1+e^{\frac{2 \alpha (L+x_0)}{k}}\right) \alpha \right) \alpha}{4 k^2 \left(k^2-h\right)}.
 \end{align*}  
For $x < x_0$, the potential $u_I(x,0)$ vanishes and equation (\ref{psixpartbox}) is again given by (\ref{psixpartsimple}).
We deduce that the second column of $\psi(x,k)$ has the form
\begin{equation}\label{psinear0}  
  [\psi(x,k)]_2 = \begin{pmatrix} b(k) e^{ikx} \\  a(k) e^{-ikx} \end{pmatrix}, \qquad x < x_0,
\end{equation}
where $a(k)$ and $b(k)$ are two coefficient functions to be determined. As the notation suggests, these coefficient functions coincide with the spectral functions $a(k)$ and $b(k)$. Indeed, (\ref{psimu3box}) and (\ref{psinear0}) imply that the eigenfunction $\mu_3(x,0,k)$ is given for $ x < x_0$ by
\begin{equation}\label{mu3x0kbox}  
  [\mu_3(x,0,k)]_2 = \begin{pmatrix} b(k) e^{2ikx} \\  a(k) \end{pmatrix},
\end{equation}
and evaluation at $x = 0$ shows that $[\mu_3(0,0,k)]_2 = (b(k), a(k))^T$ as required.

\begin{figure}
\begin{center}
\begin{tabular}{ccc}
$I = \{h, L, x_0\}$ &  \begin{tabular}{cc} Initial profile $\beta_I(x)$ \qquad \qquad & \qquad \quad  $|\Delta(k)|$  \qquad \quad \end{tabular} & $\sharp$ of solitons
	\\ \hline
	$\begin{array}{l} h = -1 \\ L = 1 \\ x_0 = 0.1 \end{array}$ 
	& \begin{tabular}{cc}
     \includegraphics[width=.3\textwidth]{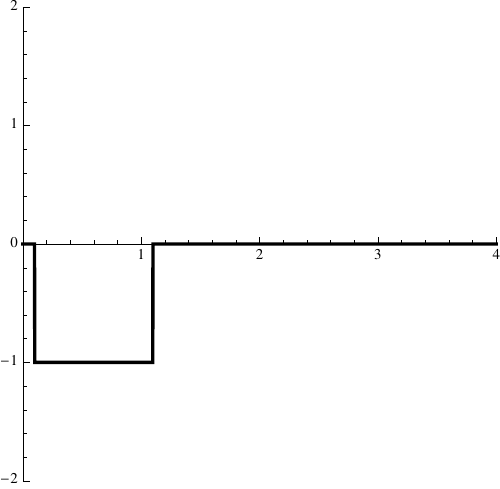} \quad
   \includegraphics[width=.3\textwidth]{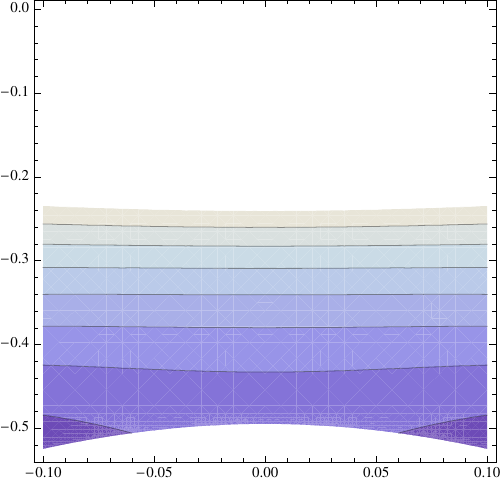} \quad
   \end{tabular}
   & 0
   	\\
	$\begin{array}{l} h = -1 \\ L = 1 \\ x_0 = 2 \end{array}$ 
	& \begin{tabular}{cc}
     \includegraphics[width=.3\textwidth]{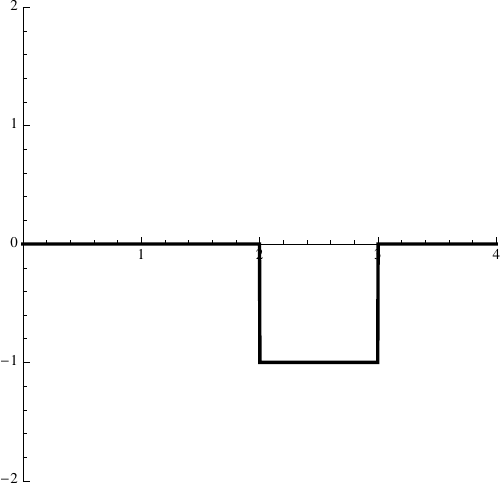} \quad
   \includegraphics[width=.3\textwidth]{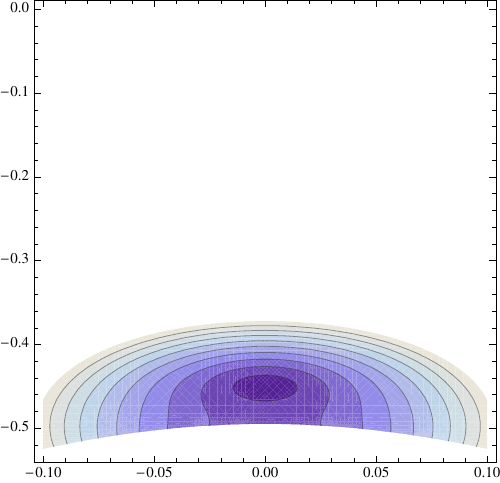} \quad
   \end{tabular}
   & $\begin{array}{c} 1 \\ _{\begin{pmatrix} p_1 = -0.449 \end{pmatrix}} \end{array}$
\end{tabular}
     \begin{figuretext}\label{box.pdf} \upshape
        The function $\Delta(k)$ for the IBV problem (\ref{boxIBVintro}) with initial profile $\beta_I(x)$ a box of height $h = -1$ and length $L = 1$ has one zero when the box lies some distance away from the boundary ($x_0 \gtrsim 1.44$), but no zeros when the box lies near the boundary ($0< x_0 \lesssim 1.44$). Hence the boundary has a direct effect on the number of asymptotic solitons.
       \end{figuretext}
   \end{center}
\end{figure}

By enforcing continuity of $\psi(x,k)$ at $x = x_0$, we deduce the following expressions for $a(k)$ and $b(k)$:
\begin{align} \nonumber
 b(k) = & \frac{h (i \cos (k (L+2 x_0))+\sin (k (L+2 x_0))) \sinh
   \left(\frac{\alpha(k) L}{k}\right)}{2 \alpha(k)},
  	\\  \label{abbox}
 a(k) = & \frac{e^{-i k L} \left(2 \alpha(k) \cosh
   \left(\frac{\alpha(k) L}{k}\right)-i \left(h-2 k^2\right) \sinh
   \left(\frac{\alpha(k) L}{k}\right)\right)}{2 \alpha(k)},
\end{align}
where $\alpha(k) =  \sqrt{h k^2-k^4}$.
Note that since the initial profile $u_I(x,0)$ vanishes for $x < 0$, the spectral functions $a(k)$ and $b(k)$ coincide with the spectral functions $a^L(k)$ and $b^L(k)$ for the problem on the line; this can also be seen directly from (\ref{mu3x0kbox}) together with the formula
$$s^L(k) = \lim_{x \to -\infty}e^{-ikx \hat{\sigma}_3}\mu_3(x,0,k).$$
In accordance with the general theory of section \ref{spectralsec}, $a(k) \to 1$ and $b(k) \to 0$ as $k \to \infty$ in $D_1 \cup D_2$, and
$$a(k) = -\frac{i \sqrt{h} \sinh \left(\sqrt{h} L\right)}{2 k} + O(1), \qquad b(k) = \frac{i \sqrt{h} \sinh \left(\sqrt{h} L\right)}{2 k} + O(1), \qquad k \to 0.$$

\subsection{Asymptotic behavior}\label{asymptoticbehaviorsubsec}
Using the expressions (\ref{abbox}) for $a(k)$ and $b(k)$, we obtain an explicit formula for the function $\Delta(k)$ defined by (\ref{aDeltadef}). The zeros of $\Delta(k)$ generate the asymptotic solitons. 
Observe that the function $\Delta(k)$ depends on $x_0$ via the spectral function $b(k)$, so that the position of the box {\it can} affect the position and the number of zeros of $\Delta(k)$. 
On the other hand, the asymptotic solitons for the infinite line problem with the same box-shaped initial profile are generated by the zeros of $a^L(k) = a(k)$. Since $a(k)$ is independent of $x_0$, its zero set is {\it independent} of the position of the box.

In order to see the effect of the boundary explicitly, we consider a box of height $h = -1$ and length $L = 1$, see Figure \ref{box.pdf}. In this case we numerically find the following:
\begin{itemize}
\item $a^L(k)$ has exactly one zero at $k \approx -0.4351 i$ independently of the value of $x_0$. 

\item For $0 < x_0 \lesssim 1.44$, $\Delta(k)$ has no zeros. 

\item For $1.44 \lesssim x_0$, $\Delta(k)$ has one zero. This zero enters $D_2$ from below at $k = - i/2$ as $x_0$ increases above $x_0 \approx 1.44$. As $x_0$ grows larger, the position of the zero moves upward along the imaginary axis. For $x_0 = 2$ the zero lies at $k \approx -0.449 i$. In the limit $x_0 \to \infty$, the zero approaches the zero of $a^L(k)$.
\end{itemize}

These observations can be understood in terms of asymptotic solitons as follows:
\begin{itemize}
\item Equation (\ref{onesolarea}) implies that the soliton $u_p^-(x,t)$ has total area $\int_{-\infty}^\infty u_p^-(x, t) dx = 4p.$
For the problem on the line there is one asymptotic soliton of total area approximately $-4 \times (0.4351) = -1.7404$. This area is comparable to the area of the box which is $-1$, and the difference can be explained by the presence of dispersive waves of predominantly positive height. The independence of the value of $x_0$ is a direct consequence of translation invariance.

\item For the half-line problem, a small value of $x_0$ means that the initial profile is a box located very near the boundary at $x=0$. The boundary condition forces the solution to vanish at $x=0$ and this has a suppressing effect on the solution near the boundary, hence no solitons have the opportunity to form. 

\item For $1.44 \lesssim x_0$, the initial profile is a box located some distance away from the boundary. The impact of the vanishing boundary condition at $x = 0$ is weak and the box is able to send away a soliton. As $x_0$ grows larger, the impact of the boundary decreases. In the limit $x_0 \to \infty$, its impact is negligible and the result on the line is recovered.
\end{itemize}

\begin{figure}
\begin{center}
\begin{tabular}{ccc}
$I = \{h, L, x_0\}$ & \begin{tabular}{ll} Initial profile $\beta_I(x)$\qquad \quad& \qquad\quad $|e^{ik L} \Delta(k)|$ \qquad \quad \end{tabular} & $\sharp$ of solitons
	\\ \hline
	$\begin{array}{l} h = -0.15 \\ L = 10 \\ x_0 = 10^{-6} \end{array}$ 
	& \begin{tabular}{cc}
     \includegraphics[width=.3\textwidth]{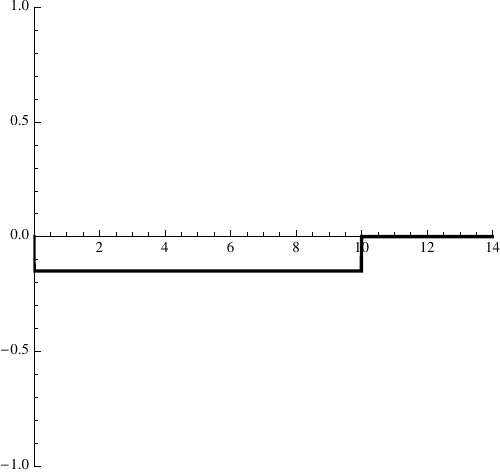} \quad
   \includegraphics[width=.3\textwidth]{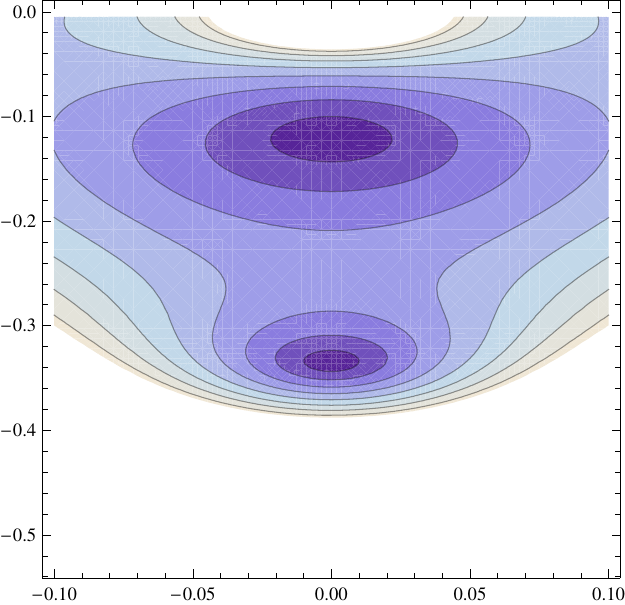} \quad
   \end{tabular}
   & $\begin{array}{c} 2 \\ _{\begin{pmatrix} p_1 = -0.335 \\ p_2 = -0.120 \end{pmatrix}} \end{array}$
   	\\
	$\begin{array}{l} h = -0.15 \\ L = 10 \\ x_0 = 2 \end{array}$ 
	& \begin{tabular}{cc}
     \includegraphics[width=.3\textwidth]{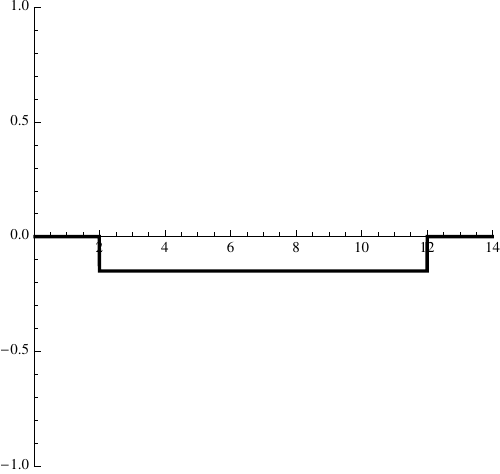} \quad
   \includegraphics[width=.3\textwidth]{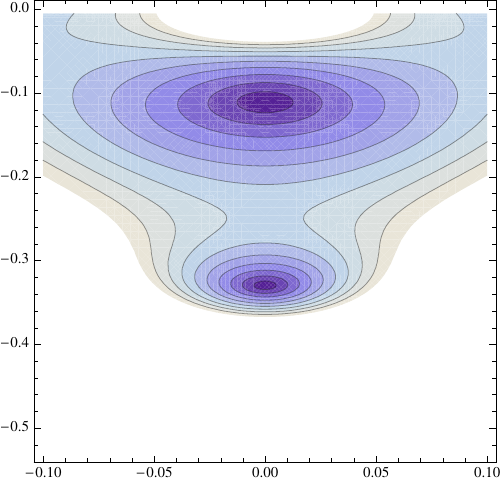} \quad
   \end{tabular}
   & $\begin{array}{c} 2 \\ _{\begin{pmatrix} p_1 = -0.330 \\ p_2 = -0.110 \end{pmatrix}} \end{array}$
\end{tabular}
     \begin{figuretext}\label{widebox.pdf} \upshape
        For a wide and shallow initial profile $\beta_I(x)$ the effect of the boundary on the asymptotic solitons is very small. The zeros of the function $\Delta(k)$ for a box-shaped initial profile of height $h = -0.15$ and length $L = 10$ are affected only very slightly by the position $x_0$ of the left edge. The contour plot of $|\Delta(k)|$ has been weighted by an exponential factor $e^{ik L}$ for clarity.
       \end{figuretext}
   \end{center}
\end{figure}

This interpretation is reinforced if we repeat the same experiment with the wide and shallow box of height $h = -0.15$ and length $L = 10$, see Figure \ref{widebox.pdf}. In this case we numerically find the following:
\begin{itemize}
\item $a^L(k)$ has two zeros at $k_1 \approx -0.330 i$ and $k_2 \approx -0.110 i$ indepedently of the value of $x_0$. 

\item In the limit $x_0 \to \infty$, on the half-line, the function $\Delta(k)$ has the same zeros as $a^L(k)$. For $x_0 = 2$ the zeros lie within $0.001$ of their limiting values.

\item Changing $x_0$ has essentially no effect on positions of the zeros of $\Delta(k)$. For $x_0 = 10^{-6}$ the zeros lie at $k_1 \approx -0.335 i$ and $k_2 \approx -0.0120 i$.

\end{itemize}

\begin{figure}
\begin{center}
\begin{tabular}{ccc}
$I = \{h, L, x_0\}$ & \begin{tabular}{cc} Initial profile $\beta_I(x)$ \qquad \qquad & \qquad \quad  $|\Delta(k)|$  \qquad \quad \end{tabular} & $\sharp$ of solitons
	\\ \hline
	$\begin{array}{c} h = 1 \\ L = 1 \\ x_0 = 1 \end{array}$ 
	& \begin{tabular}{cc}
     \includegraphics[width=.3\textwidth]{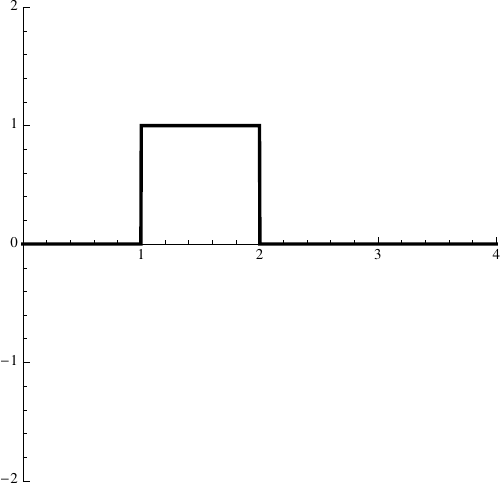} \quad
   \includegraphics[width=.3\textwidth]{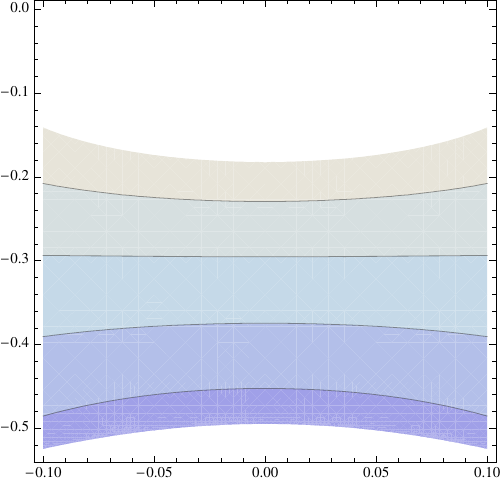} \quad
   \end{tabular}
   & 0
\end{tabular}
     \begin{figuretext}\label{boxhpositive.pdf} \upshape
        For a box-shaped initial profile $\beta_I(x)$ of positive height $(h > 0)$ there are no solitons generated asymptotically. For the half-line problem this follows from the absence of zeros for $\Delta(k)$.
       \end{figuretext}
   \end{center}
\end{figure}

\begin{remark}\upshape
The interpretations of these observations are similar to those for the box with $h = -1$ and $L = 1$ above, except that now the box is very wide and shallow and since the effect of the boundary is only felt at the left edge, the impact of the boundary condition is substantially reduced.

As a final conclusion drawn from the numerical observations, we mention that whenever $h > 0$, there appear to be no asymptotic solitons, neither on the half-line nor on the line, see Figure \ref{boxhpositive.pdf}. This observation is in accordance with the fact that the right-traveling solitons are waves of depression (see equation (\ref{onesoliton})) and cannot be generated by a box of positive height. 
\end{remark}

\appendix
\section{Scalar Lax pair} \label{scalarlaxapp}
\renewcommand{\theequation}{A.\arabic{equation}}\nequation
Although it is usually more convenient to use the matrix Lax pair presented in (\ref{lax}), equation (\ref{kdv}) can also be formulated as the condition of compatibility of the following scalar Lax pair:
\begin{equation}\label{scalarlax}
\begin{cases}
  & \varphi_{xx} = (u + k^2)\varphi,
  	\\
  & \varphi_t = u_x\varphi + (4k^2 - 2q - 1)\varphi_x,
  \end{cases}
\end{equation}
where $k \in \C$ is a spectral parameter and is a constant and $\varphi(x,t,k)$ is an eigenfunction. The scalar and matrix Lax pairs are related as follows. Let $u(x,t)$ be given and let $\varphi(x,t,k)$ be the solution of (\ref{scalarlax}) satisfying $\varphi(x,t,k) \sim e^{kx}$ as $x\to \infty$. Let $\mu_3(x,t,k)$ be the matrix eigenfunction defined in (\ref{mujdef}) satisfying $\mu_3(x,t,k) \to I$ as $x \to \infty$. Then
\begin{equation}\label{varphitomu3}
\mu_3(x,t,k) =  \frac{1}{2}\begin{pmatrix} \varphi(x, t, ik) - \frac{i}{k}\varphi_x(x, t, ik)	&	\varphi(x, t, -ik) - \frac{i}{k}\varphi_x(x, t, -ik) \\
\varphi(x, t, ik) + \frac{i}{k}\varphi_x(x, t, ik)	&	\varphi(x, t, -ik) + \frac{i}{k}\varphi_x(x, t, -ik) \end{pmatrix}
e^{i(-kx + (k + 4 k^3)t)\sigma_3}.
\end{equation}

 \bigskip
\noindent
{\bf Acknowledgement} {\it The authors acknowledge support from a Marie Curie Intra-European Fellowship, EPSRC, and the Guggenheim foundation, USA.}

\bibliography{is}

\end{document}